\documentstyle[aaspptwo]{article}
\newcommand{\etal}{{{ et al.}}~}
\newcommand{\eg}{{{e.g.,}}~}
\newcommand{\ie}{{{i.e.,}}~}
\newcommand{\kms}{{{km s$^{-1}$}}~}

\newcommand{\kmsp}{{{km s$^{-1}.$}}~}
\newcommand{\Ho}{{{H$_\circ$}}~}
\newcommand{\Hop}{{{H$_\circ.$}}~}
\newcommand{\Hoc}{{{H$_\circ,$}}~}
\newcommand{\amin}{{{$^\prime$}}~}

\slugcomment{Submitted to the Astrophysical Journal}

\begin{document}

\title{Systematic Errors in the Hubble Constant Based on
Measurement of the Sunyaev-Zeldovich Effect}

\vspace{1.in}
\author{KURT ROETTIGER$^1$, JAMES M. STONE,}
\affil{Department of Astronomy \\ University of Maryland\\ College Park, MD 20742-2421\\
e-mail: kroett@astro.umd.edu\\jstone@astro.umd.edu}
\author{RICHARD F. MUSHOTZKY}
\affil{Goddard Space Flight Center\\Code 662.0\\Greenbelt, Maryland 20771\\
e-mail: mushotzky@lheavx.gsfc.nasa.gov}

\vspace{.5in}

\begin{center}{\bf To  appear in the Astrophysical Journal June 20, 1997}
\end{center}

% Notice that each of these authors has alternate affiliations, which
% are identified by the \altaffilmark after each name.  The actual alternate
% affiliation information is typeset in footnotes at the bottom of the
% first page, and the text itself is specified in \altaffiltext commands.
% There is a separate \altaffiltext for each alternate affiliation
% indicated above.

\altaffiltext{1}{Current Address: Goddard Space Flight Center, Code 930, Greenbelt, MD 20771}. 
%CTIO is operated by AURA, Inc.\ under contract to the National Science
%Foundation.} 
%\altaffiltext{2}{Society of Fellows, Harvard University.} 
%\altaffiltext{3}{present address: Center for Astrophysics,
%    60 Garden Street, Cambridge, MA 02138}
%\altaffiltext{4}{Visiting Programmer, Space Telescope Science Institute}
%\altaffiltext{5}{Patron, Alonso's Bar and Grill}

% The abstract environment prints out the receipt and acceptance dates
% if they are relevant for the journal style.  For the aasms style, they
% will print out as horizontal rules for the editorial staff to type
% on, so long as the author does not include \received and \accepted
% commands.  This should not be done, since \received and \accepted dates
% are not known to the author.

\begin{abstract}

Values of the Hubble constant reported to date which are based on measurement of the
Sunyaev-Zeldovich (SZ) effect in clusters of galaxies are systematically lower than those
derived by other methods (e.g., Cepheid variable stars, or the
Tully-Fisher relation).  We investigate the possibility that systematic errors
may be introduced into the analysis by the generally adopted
assumptions that observed clusters are in hydrostatic equilibrium, are
spherically symmetric, and are isothermal. We construct self-consistent theoretical
models of merging clusters of galaxies using hydrodynamical/N-body simulations. We then
compute the magnitude of \Ho derived from the SZ effect at different
times and at different projection angles both from first principles,
and by applying each of the standard assumptions used in the interpretation of observations.
Our results indicate that the assumption of isothermality in the evolving clusters can result in
\Ho being underestimated by 10-30\% depending on both epoch and projection angle. 
Moreover, use of the projected, 
emission-weighted temperature profile under the assumption of spherical symmetry does not
significantly improve the situation except in the case of more extreme mergers (\ie those
involving relatively gas-rich subclusters). Although less significant, we find that asphericity 
in the gas density can also result in a 15\% error in \Hop
If the cluster is prolate (as is generally the case for on-axis, or nearly on-axis
mergers), and viewed along
its major axis, \Ho will be systematically underestimated. More extreme offaxis mergers
may result in oblate merger remnants which when viewed nearly face-on may result in
an overestimation of \Hop A similar effect is noted when viewing a prolate distribution
along a line-of-sight which is nearly perpendicular to its major axis. 
In both cases the potential overestimation occurs only
when the remnant is viewed within 15-30$^\circ$ of face-on.
 Bulk gas motions and the
kinematic SZ effect do not appear to be
significant except for a brief period during the very early stages of a merger.  Our study
shows that the most meaningful SZ measurement will be accompanied by a high resolution
temperature data and a detailed dynamical modeling of the observed system. In lieu of this,
a large sample selected to avoid dynamically evolving systems is preferred.

\end{abstract}


\section { INTRODUCTION}
The Hubble constant (\Ho) and determination of its value are among the most
controversial topics within the astronomical community.
Recently, however, there seems to be a general convergence toward a value for \Ho
at the high-end of the currently accepted range, (\ie $\ge$ 70
km s$^{-1}$ Mpc$^{-1}$). Such a value is supported by a variety of methods including
the Tully-Fisher relation (85$\pm$10; Pierce \& Tully 1988), galaxy surface brightness fluctuations 
(82$\pm$7; Tonry 1991), Type II supernovae and the expanding photosphere method (EPM) (73$\pm$7; Schmidt
\etal 1994), as well as both space-based (80$\pm$17; Freedman \etal 1994) and ground-based
(87$\pm$7; Pierce \etal 1994) measurements of Cepheid variable stars in the Virgo cluster. 
Mould \etal (1995) provides an excellent review of the recent analysis
applied to the Virgo cluster in which they note that all of the above numbers are consistent with 80$\pm$17
km s$^{-1}$ Mpc$^{-1}$.

In contrast to these values, two methods have consistently supported a lower ($<$ 60 km s$^{-1}$ Mpc$^{-1}$)
value for \Hop Schaefer (1996) reviews the observations of 10 Type Ia supernovae (SNIa) and
reports a mean value of 55$\pm$3. However, Hamuy \etal (1995) finds 
that values of \Ho based
on SNIa will be systematically underestimated by 15\% unless a correction is applied using
the luminosity-decline rate relation of Phillips (1993) which leads to a best value
near 65 km s$^{-1}$ Mpc$^{-1}$. This result is consistent  with the SNIa-based value derived
by Riess, Press, \& Kirshner (1994), 67$\pm$7 km s$^{-1}$ Mpc$^{-1}$.   Mould \etal (1995), using
a recalibration of SNIa absolute magnitudes based on six  SNIa in the Virgo cluster, find a value
of 71$\pm$7 km s$^{-1}$ Mpc$^{-1}$.

The second method which has consistently supported lower values of \Hoc and 
the one of interest here, relies on
the measurement of the Sunyaev-Zeldovich (SZ) effect (Sunyaev \& Zeldovich 1972).
For a complete and recent review of the SZ effect, the reader is referred to Rephaeli (1995).
Briefly, the SZ effect is a distortion of the cosmic microwave background (CMB) spectrum 
caused by inverse Compton scattering of low
energy CMB photons by high energy electrons in the dense cores
of massive clusters of galaxies. Knowing the density and temperature of the scattering
plasma, as determined from X-ray observations of the intracluster medium (ICM), and the
CMB brightness temperature in the cluster core, as measured by millimeter observations, it
is possible to determine \Ho (see \S \ref{sz}). Owing to the difficulty of the millimeter observations 
and, until recently, the limited quality of X-ray temperature data, there have only been
a few determinations of \Ho using this method. Still, they have tended to support
a lower value for \Ho (\eg  45$\pm$17, Birkinshaw \etal 1991; 
48$\pm$28, Jones \etal 1993; 55$\pm$17 Birkenshaw \& Hughes 1994). The one notable exception
is a recent measurement using the Coma cluster which finds \Ho= 71$\pm^{30}_{25}$ 
km s$^{-1}$ Mpc$^{-1}$ (Herbig \etal 1995). This value is based on a Ginga temperature
for Coma of 9.1 keV. New  ASCA observations resulted in a
central temperature for Coma of $\sim$8 keV (Honda \etal 1996) which is more consistent with
previous temperature estimates such as 7.5$\pm$0.2 keV from {\it Tenma}, 8.5$\pm$0.3 keV from 
{\it EXOSAT} (Hughes 1989),  and 7.7 keV  from {\it Spacelab 2} (Watt \etal 1992).
Using the lower temperature estimate would reduce the the Coma-derived value
of \Ho to a value nearer 50 km s$^{-1}$ Mpc$^{-1}$. The review article by Rephaeli (1995)
lists two other measurements (A2256 and A2163) of \Ho using the SZ effect which are greater than 
70 km s$^{-1}$ Mpc$^{-1}$.
However, both of these cluster show significant substructure, neither of these \Ho measurements 
have appeared, as of yet, in the refereed journals.  Similarly,
Kobayashi, Sasaki \& Suto (1996) argue that SZ based measurements are consistent with those
of Cepheids. This consistency is owed to the large uncertainty in the SZ derived values of \Hop

What are the sources of this apparent systematic discrepancy?  Although the X-ray data have 
improved dramatically in the last decade,
it is still difficult to determine the internal structure of clusters from the X-ray imaging
because it only supplies projected temperature and surface brightness 
information, while supplying no information regarding the internal gas dynamics. Consequently, observers 
are often forced to assume that clusters are spherical, hydrostatic, and isothermal systems.
Such assumptions are contrary to recent observational evidence that shows many clusters
(30-70\%, Jones \& Forman 1991; Bird 1993; Davis 1994; Mohr \etal 1995) are still dynamically evolving.
In a few instances, detailed temperature maps of clusters have been produced only to reveal
extremely complex temperature morphologies (\eg 2256,
Briel \& Henry 1994; Roettiger, Burns \& Pinkney 1995;  A754, Henriksen \& Markevitch 1996) 
 We attempt to determine the source of the discrepancy
using self-consistent numerical hydrodynamical/N-body simulations of evolving clusters of galaxies.
Specifically, we address systematic errors associated with the inherent limitations
in the X-ray data.
Our basic approach is to ``observe"  massive simulated
clusters at several epochs during mergers with a variety of subclusters. We focus on mergers, because
it is believed that massive clusters grow through the accretion of smaller structures, and mergers are the only 
mechanism likely to affect a cluster's global properties. Using the same techniques
commonly employed by observers, we determine the temperature and density profiles of the simulated
clusters as a function of viewing angle and epoch. We then compare the ``observed" value
of \Ho with the ``true" value calculated from the line-of-sight (LOS) temperature
and density profiles, \ie from first principles. By comparison of the observed and true values of
\Hoc we can evaluate the systematic errors.

It should be emphasized that we are analyzing the remnants of strong mergers which are therefore disrupted
systems, and in some sense, they may represent a worst case scenario. On the other hand,
it should be kept in mind that selection effects exist such that a sample of clusters having
SZ measurements can potentially be biased toward evolving systems such as these. We find
that such systems are hotter and, consequently, more X-ray luminous then their relaxed counterparts.
Both effects will bias their selection. In addition, we find merger remnants to be prolate systems.
Consequently, they will have an enhanced X-ray surface brightness when viewed end-on  which 
may further bias
their selection. It should be noted that all but two of the clusters
listed in the sample presented by Rephaeli (1995) contain significant
substructure, and those two (A478 and A2142) are considered to be
strong cooling flows (Edge \etal 1992), a matter that further
complicates
the SZ analysis.

In \S \ref{sz}, we describe the Sunyaev-Zeldovich effect and how it is used to
determine \Hop Section \ref{num} describes our numerical methodology and
initial conditions for computing cluster mergers, as well as, our choice of parameter
space, our data analysis, and tests of our method. 
 Section \ref{evolve} contains a physical description
of the cluster mergers. Section \ref{results} contains our major
results and a comparison with results from a previous study by Inagaki \etal (1995). 
In \S \ref{discus}, we discuss the origin of the observed discrepancy, limitations
of this study, and implications for observers. And, finally,  we summarize
our conclusions in \S \ref{conclusions}

\section {THE SUNYAEV-ZELDOVICH EFFECT AND THE HUBBLE CONSTANT}
\label{sz}
The SZ effect is a spectral distortion of the cosmic microwave
background (CMB) resulting from inverse Compton scattering of low energy
CMB photons by high energy electrons in the ICM  (Sunyaev \& Zeldovich 1972). The change in the
CMB brightness temperature observed in the Rayleigh-Jeans region of the spectrum
is:
\begin{equation}
\label{dt}
\frac {\Delta T} {T_{CMB}} = -2 \frac {\sigma_T k} {m_e c^2} \int n_e T_e dl,
\end {equation}
where $n_e$ is the electron density, $T_e$ is the electron temperature, $\sigma_T$ is 
the cross section for Thomson scattering, and
the line integral, $dl$ , is performed along the LOS through the cluster. 
Since $n_e \propto h^{1/2}$ and $dl \propto h^{-1}$ (T$_e$ is independent of $h$), we can
solve Eq. \ref{dt} for $h$ in terms of the observable quantities $\Delta T$ 
(millimeter observations), $n_e(r)$ and $T_e(r)$ (X-ray observations),
\begin{equation} 
\label{heqn}
  h \propto \left ( \frac { \int n_e T_e dl} { \Delta T/ T_{CMB}} \right) ^2.   
\end{equation}
A more detailed derivation of the dependence of \Ho on these quantities
can be found in Birkinshaw \etal (1991) and Rephaeli (1995).  Rephaeli (1995)
also includes the relativistically correct form.

Observationally, the situation is complicated by several factors. Typical values for
the temperature decrement are of order a few hundred $\mu$K thus requiring
extreme sensitivity of the millimeter measurements which in turn limits this type
of observation to the hottest, densest, and presumably most massive clusters. Often, these are distant clusters
which are  affected by low spatial resolution in both the millimeter and X-ray observations.
Although low spatial resolution and the use of a reference beam in the millimeter
observations introduces a significant model dependence to the analysis, the central averaging
may actually help smooth small scale temperature fluctuations and thereby help mitigate
some of the associated uncertainties (Herbig 1996).
The typical resolution of the millimeter observations are $>$1\amin (\eg 1.78\amin, 
Birkinshaw \& Hughes 1994; 7.3\amin, Herbig \etal 1995). 
It is therefore necessary to make  assumptions about the  cluster gas distribution in 
order to deduce the central temperature
decrement from the observed beam averaged value (Birkinshaw \etal 1984).  Only recently have interferometric
techniques allowed the generation of SZ maps with sub-arcmin resolution (\eg Carlstrom \etal 1996).
In addition to these sources of error, observers must contend with contamination by cluster radio sources.

In this study, we do {\it not} address the uncertainties associated with the measurement of $\Delta T/T$.
Therefore, from Eq. \ref{heqn}, the fractional error in \Ho based
solely on limitations of the X-ray data is simply the square of the ratio of the observationally derived
pressure integral to the true LOS pressure integral or,
\begin{equation}
\label{ratio}
\frac {h_{obs}} {h_{true}} = \left [ \frac {\left [ \int n_e T_e dl \right ]_{obs} } 
{\left [\int n_e T_e dl \right ]_{true}  } \right ]^2
\end{equation}
where the RHS numerator is  based on observationally
derived values for $n_e$, $T_e$, and $dl$ and the denominator is based on the
true 3-dimensional simulation data. Both integrals are calculated at the resolution of the simulation, 50 kpc
(See \S \ref{num}).

\section {METHOD}
\label{num}
\subsection {Dynamical Calculations of Cluster Mergers}
\subsubsection {Numerical Method}

To compute the dynamical evolution of merging clusters of galaxies, we use a hybrid hydrodynamical/N-body code in which the hydrodynamical component is
CMHOG written by one of us, J. M. Stone.  CMHOG solves the fluid equations using an implementation of the 
piecewise-parabolic method (PPM, Colella \& Woodward 1984) in its Lagrangian remap formulation for
gas dynamics. The primary advantages of PPM are lower numerical diffusion through the use
of third order parabolic spatial interpolations, and
its ability to accurately resolve shocks in as few as 1-2 zones through the use of
a Riemann solver. The result is an increase in the small structure captured at a given
resolution over codes employing artificial viscosity to capture shocks. This code has been applied to a 
variety of astrophysical 
problems and has been extensively tested using the problems described in Woodward and Colella (1984). 

The collisionless dark matter is evolved using an N-body code based on a standard 
particle-mesh algorithm (PM). The particles
are evolved on the same grid as the gas using the same time step. The time step
is determined by applying the Courant condition simultaneously to both the dark matter and the hydrodynamics.
The only interaction between the collisionless particles and the gas is via the
gravitational potential. Therefore, the
two codes are run largely independent of each  other. The hydrocode supplies
the gas density distribution to the N-body code as part of the source function to Poisson's
equation (the self-gravity of both the gas and dark matter is included in these simulations). 
The N-body code, in turn, supplies the gravitational potential to the hydrocode
where it becomes a source function to the energy equation. Since we are modeling an isolated
region, the boundary conditions for Poisson's equation are determined by a multipole expansion
of the mass distribution contained within the grid.  Particles that leave the grid are
lost to the simulation. Typically, less than a few percent of the particles leave the grid.

The computational grid is similar to that described in Roettiger, Loken, \& Burns (1997). The
simulation is fully three-dimensional. The grid is a fixed and rectangular (100 x 100 x 220 zones).
The merger axis coincides with the grid's major axis along which 
 resolution is uniform and scales to 50 kpc or 6 zones per
primary cluster core radius. Resolution along the grid's minor axis is uniform within the central 
40 zones and ratioed in the 30 zones on either side. That is, the resolution
is a uniform 50 kpc extending to 1 Mpc (20 zones) on either side of the merger axis. Beyond 1 Mpc,
the zone dimensions increase by 3\% from one zone to the next out to the edge of the grid.
We use outflow boundary conditions for the hydrodynamical evolution.  We do not include
radiative cooling in these simulations.

\subsubsection {Initial Conditions}

The initial conditions are similar to those in several previous studies (Roettiger, Loken \&
Burns 1993; Roettiger, Burns \& Pinkney 1995; Roettiger \etal 1996, 1997). We begin with two clusters that are initially isothermal ($r < 4r_c$) and in
hydrostatic equilibrium. They are simply placed on the computational grid separated by 
$\sim$ 5 Mpc and allowed to merge under the influence of their mutual gravity. The clusters
are given an initial relative velocity of 300 \kmsp  While not significantly affecting the final impact
velocity, the small initial velocity does greatly reduce the time required to merge and thus
saves considerable computational resources.
The idealized initial conditions have several benefits. They allow a more efficient
use of the computational volume, and consequently, higher resolution.  They
provide a well-defined baseline with which to compare the subsequent evolution
of the post-merger clusters. And, they allow for a systematic study of merger parameter
space. 

Whereas the basic setup is identical to previous simulations by Roettiger \etal, the
details of the intial cluster structure have changed somewhat. The total
mass distribution in these simulations is well-characterized by a power-law
slope of approximately -2.5 for $r>r_c$.  The gas distribution is typically 
somewhat flatter, ranging from -2.0 to -2.5 depending on the intial $\beta$ parameter, 
see Table 1. The temperature profiles are also  somewhat different from
previous simulations. These clusters are strictly isothermal only within 4$r_c$. At larger radii,
the temperature decreases gradually. Figure \ref{initprof} shows an example of the initial
density and temperature profiles for a $\beta$=1 cluster. Here, we define two values of
$\beta$, designated $\beta_{fit}$ and $\beta_{spec}$. The value of $\beta_{fit}$ is determined
by fitting an isothermal $\beta$ model to the X-ray surface brightness distribution (see eq. 5), while
$\beta_{spec}=\mu m_p {\sigma_v}^2/ k T$ where $\mu$ is the mean molecular weight, $m_p$ is the
proton mass, $k$ is Boltzmann's constant, $\sigma_v$ is the galaxy velocity dispersion, and $T$
is the ICM temperature.  Initially, $\beta_{fit}$=$\beta_{spec}$, at least within the central
3-4$r_c$. Unless otherwise stated, all references to $\beta$ refer to $\beta_{fit}$. Finally, since
these simulations are non-cosmological, the value of \Ho does not directly influence the
scaling of the cluster dimensions. However, the arbitrary length scaling that we have chosen most
closely resembles clusters in a universe in which \Ho=50 km s$^{-1}$ Mpc$^{-1}$.

\subsubsection{Parameter Space}

In order to constrain our merger parameter space, we need to ask several questions. First,
given a massive cluster of the type necessary for the SZ analysis, what mass of cluster
is it most likely to have  recently interacted with it? We gain some insight into this 
matter by looking at the mass function for clusters. Bahcall \& Cen (1993) analytically 
represent the cluster mass function as, 
\begin{equation}
n(>M)=4 \times 10^{-5}(M/M^*)^{-1}exp(-M/M^*) h^3 Mpc^{-3}
\end{equation}
where $n(>M)$ is the number density  of clusters with mass greater than $M$,
$M^* = 1.8\pm0.3 \times 10^{14} h^{-1}$ M$_\odot$, and $h$ = H$_\circ/100$.
Normalizing to a region of space large enough to have one cluster of mass
comparable to our primary ($\sim$10$^{15}$ M$_\odot$), we find that $\sim$90\% of the 
other clusters in this same region are $\le$25\% of its mass. Thus, we focus on
mergers between clusters with large mass ratios.

A second consideration is the relative gas content of the two clusters, particularly their
 central gas densities which, when combined with the impact velocity determines
the ram pressure experienced by the respective clusters. Previous simulations have shown the 
importance of the
self-interaction of the gas components in determining the post-merger evolution of the
ICM.  The relative gas densities will determine the degree to which the subcluster
gas is stripped, the strength of shocks generated during the merger, and the accompanying
heating, as well as the size and duration of bulk flows and the rate at which the
remnant returns to equilibrium. X-ray studies of clusters of galaxies show the
typical range of central gas densities to be 10$^{-4} \le n \le$ 10$^{-2}$ cm$^{-3}$ (Sarazin 1986),
while the global gas content of clusters, by mass, seems to range from a few percent to $>$30\% 
with typical values in the 10-20\% range (White \& Fabian 1995).

Since we are modeling a wide range in total cluster mass (up to a factor of 8), it would be helpful
to constrain the models if there were a correlation between total mass and the ICM
properties.  Assuming hydrostatic equilibrium is the norm rather
than the exception, one would expect a reasonable correlation between total mass and ICM
temperature. However, a similar correlation does not appear to exist between
total mass and gas density or baryon fraction (White \& Fabian 1995).
 In the absence of some extreme local physics (\eg AGN,
star formation), one would expect the fraction of mass in baryons to be relatively constant
from one cluster to the next. The fact that we observe a wide range of values 
is, in and of itself, a very interesting problem with considerable cosmological implications 
(\eg Lubin \etal 1996).
Still, it does little to help constrain our models. The situation is no better with the
central gas density. Here, studies are limited by the resolution of the X-ray observations, and
it may be more reasonable to expect local physics, particularly radiative cooling, to play a 
significant role.  It has been noted that many very low mass systems tend to have more extended
gas distributions than their high mass counter parts (\ie low $\beta$) (Doe \etal 1995; Mulchaey \etal 1996). 
In which case, if the
gas contents are comparable and the dark matter distributions similar, one might then expect
at least on average, that lower mass systems will have lower central densities.

Table 1 contains the parameters of the clusters used in this analysis. In
light of the above considerations, we have run six head-on merger simulations in which we
vary the relative cluster masses and gas properties. In each case, the primary cluster
is scaled to 1.5 $\times$ 10$^{15}$ M$_\odot$. We have evolved both 4:1 and 8:1 total mass ratio mergers.
Within each mass ratio, we vary both the gas content and central gas densities within
the observed range of parameters. Central gas density varies from $\sim 8 \times 10^{-4}$ to nearly
2 $\times 10^{-2}$ cm$^{-3}$.  The fraction of mass in baryons within the central 1 Mpc ranges from
0.08 to 0.22 with most clusters $\sim$0.12.  Merger 7 is an offaxis merger in which the impact
parameter is $\sim$160 kpc (see \S\ref{offaxis}). Although this study in no way represents a complete sampling of parameters
space, we do believe it to be representative of that space most likely to affect the SZ
analysis.  Mergers with lower mass subclusters or  those with lower central gas densities 
relative to the primary are less likely
to significantly influence the internal structure of the primary cluster.

\subsection{Synthetic Data Analysis of Merger Simulations}
\label{analysis}
In this section we describe the manner in which we attempt to obtain the X-ray
observables, $T_e$ and $n_e$, from the simulated data base.
It is our goal to mimic the observational procedures as closely as possible
in order to examine the effects of limited information in the observational
data.  We do not address problems associated with observational concerns such
as  photon statistics or poor spectral resolution. We therefore assume
the clusters are well-observed within the region being analyzed.

\subsubsection{Observed Temperature}

Observationally, we are interested in the projected, emission-weighted ICM temperature.
After selecting a particular epoch within
a given merger, we calculate the X-ray volume
emissivity ($\varepsilon_V$) within a chosen bandpass (\eg ASCA, 0.5-10 keV).
We use the Mewe-Kaastra-Liedahl emission spectrum for optically-thin plasmas supplied with XSPEC.
The model includes both thermal bremsstrahlung and line emission. We have assumed the clusters 
to have 0.3 solar 
abundance (Mushotzky \etal 1996).  We then
weight the electron temperature within a given zone (which, of course, is
given directly by the hydrodynamical evolution) by the volume emissivity within that 
same zone. The resulting product, $\varepsilon_V  T_e$, is then integrated along each LOS through
the simulation volume for a chosen projection angle creating an image which is
then divided by the total emissivity along each LOS. The result is a projected, emissivity-weighted
temperature map of the cluster at approximately the same resolution of the
simulation itself (\ie 50 kpc). The isothermal temperatures used in the following analysis
are determined by averaging the above temperature map within a given aperture
centered on the emission peak.  In \S \ref{results}, we will explore the effect of varying aperture
size. Similarly, the temperature profiles are
generated by an azimuthal averaging of the temperature map within concentric annuli
centered on the emission peak. In both cases, it is important to note that in the
absence of spherical symmetry, temperature structure along the LOS is lost.

\subsubsection{Observed Density Profile}

The procedure used to determine the effective electron density measured
by observers, $n_e(r)$, is somewhat more complicated, and we simulate the procedure
used by most observers. As before,
 $\varepsilon_V$ is calculated at a given epoch. A LOS integration through
the simulation volume at a chosen projection angle results in an  X-ray surface
brightness image.  The X-ray surface brightness profile, $S_x(r)$ , is generated by an azimuthal
averaging of the surface brightness image within annular rings centered on
the peak surface brightness. The resulting profile
is then fit with an isothermal $\beta$-model (Sarazin, 1986),
\begin{equation}
\label{sx}
S_x(r)=S_o \left [ 1 + \left (\frac {r} {r_c} \right ) ^2 \right ]^{-3\beta + \frac {1} {2}},
\end{equation}
where the core radius, $r_c$, and $\beta$, are free parameters.
We choose 1 Mpc as a reasonable extent for the observable X-ray emission.
It is roughly three times the initial core radius. Generally,
one can not expect a good fit to the $\beta$-model much beyond 2-3 core radii. It should be
noted that the fit parameters, $r_c$ and $\beta$, can depend strongly on
the maximum radius used in the fit, and that this will contribute to some of the
error in \Hop However, in general, $r_c$ and $\beta$ are
not fit independently and vary such that the line integral of the $\beta$-model remains somewhat constant.
This is important since the SZ effect depends on the line integral not on the individual
parameters. 

Given the fit parameters, $r_c$ and $\beta$, and the peak X-ray surface
brightness, $S_o$, it is possible to determine the central electron density, $n_{o}$,
using the following relation based on Eq. 3 in Henry \& Henriksen (1986),
\begin{equation}
S_o=N(E_1/kT_e,E_2/kT_e)\frac {\Gamma(3\beta - 1/2)} { \Gamma(3\beta)} r_c n_o^2 (kT_e)^{1/2}.
\end{equation}
Here, $N$ is a normalizing factor dependent on the bandwidth ($E_1, E_2$) of the X-ray observation,
$T_e$ is the electron temperature, $k$ is Boltzman's constant, and $\Gamma$ is the complete
gamma function.  Knowing $r_c$, $\beta$, and $n_o$, we can now calculate the electron
density profile as follows,
\begin{equation}
n_e(r)=n_o \left [ 1 + \left (\frac {r} {r_c} \right ) ^2 \right ]^{-3\beta/2}.
\end{equation}
As with the temperature data, in the absence of spherical symmetry, all LOS density
inhomogeneities are lost. Furthermore, the perceived shape of the density distribution ($r_c$, $\beta$, and
ellipticity) may differ significantly from the true distribution. 

Since we do not include errors associated with photon statistics, all uncertainty
in the fit parameters and ultimately in the electron density result from inhomogeneities and asphericity
in both the density and temperature distributions or from a real deviation from a King model. For consistency in the analysis, 
the X-ray profile is always centered on the highest peak in the X-ray emission. The ``best"
fit parameters are determined objectively by a $\chi^2$ minimization
routine.

\subsection {An Isolated Cluster: A Test of the Model}
\label{single}

The above analysis was applied to a single, hydrostatic cluster which is similar to
the primary cluster in each merger.  The test cluster was placed on the same
grid used in all the merger simulations only now it was allowed to evolve in isolation
for a time period comparable to that of the mergers, $\sim$8 Gyrs. This test
serves several purposes. First, it demonstrates the degree of dynamical stability in our model
cluster. The cluster begins isothermal within $\sim$4$r_c$ and in hydrostatic
equilibrium. Left on its own, it should remain so. However, due to resolution
effects (largely in the N-body distribution), and infalling ambient matter, we do expect
some minor evolution of the cluster. Second, we can test the observational analysis
described above (\S \ref{analysis}) by attempting to reproduce the temperature and gas
density profiles of the cluster. Finally (and most importantly), in light of the inevitable, 
albeit minor, evolution of the isolated cluster and the limited observational 
information, we can attempt to accurately reproduce \Ho using the SZ effect. This final test  allows us to
assess the degree to which the observed \Ho discrepancy is attributable
to recent merger activity and the evolving dynamical state of the clusters.

The most noticeable evolution occurs in the ICM
temperature. Initially, the cluster is isothermal (6.5 keV) within the central
1.2 Mpc or 4 core radii. After 4 Gyrs, the time at which core passage 
occurs in the merger simulations, the temperature within this same region has risen to 6.9 keV with a
zone-to-zone dispersion of 0.22 keV or roughly 3\% of the mean. After 8 Gyrs,
the temperature has risen to nearly 7.5 keV with a dispersion of 0.29 keV or
$<$4\% of the mean.  Since the ICM heating is rather
uniform across the cluster, we believe that
the primary cause is a general, although slight, contraction of the cluster mass distribution. 
There is some evidence from the fitting of $\beta$ models
to the synthetic X-ray images that the cluster core has contracted by $\sim$5\% in
radius, or $\sim$15\% in volume. In addition, some heating undoubtedly
arises from random fluctuations in the gravitational potential, particularly in the core. This serves
to heat the gas by inducing collisions through random  motions. It is this process that
likely produces most of the dispersion about the mean temperature. In any case,
the cluster remains largely isothermal as indicated by the low temperature dispersions, and
evolution of the mean temperature is small compared to that seen in the merger simulations.
Finally, Figure \ref{sngled} shows the gas density profile at 0, 4 and 8 Gyrs. As with temperature,
the density evolution is minimal. A slight increase in the core density is noted. This is likely
related to the contraction of the cluster discussed above. The $\beta$-model fitting
indicates that the cluster remains spherical to better than a few percent.

In order to test the significance of the observed evolution to the SZ analysis, we have
performed the observational analysis described in \S \ref{analysis}  At each epoch,
we calculate the true SZ-based \Ho at 100 random LOS through the cluster core while simultaneously
calculating the observed value of \Hop We find at 0 Gyrs, $<h_{obs}/h_{true}>=1.00\pm0.004$,
at 4 Gyrs, $<h_{obs}/h_{true}>=1.00\pm0.01$, and at 8 Gyrs, $<h_{obs}/h_{true}>=1.03\pm0.02$.
In each analysis, the cluster was assumed to be isothermal.
In the last instance, we find a +3\% discrepancy in the observed value of \Hop Our
analysis shows that the low level of non-isothermality and small degree of
asphericity discussed above contribute  about equally to this systematic effect.
 We performed the same analysis using the projected, emission-weighted temperature profiles, and
found $<h_{obs}/h_{true}>=1.00\pm0.004$, 0.99$\pm$0.02, and 1.00$\pm$0.02 for 0, 4 and 8 Gyrs, respectively.
This test demonstrates that 1) our analysis procedure can accurately reproduce the temperature and
density profiles, and 2) the minimal evolution observed in the test cluster does not significantly influence the
SZ analysis, and therefore, the results described in \S \ref{results} can be attributed
largely to the post-merger evolution.

\section {A DESCRIPTION OF A TYPICAL CLUSTER-SUBCLUSTER MERGER EVENT}
\label{evolve}
A detailed description of the physical evolution of cluster-subcluster mergers can be
found in Roettiger \etal (1997). Here, we will only briefly review the evolution with an
emphasis on those aspects most pertinent to the SZ analysis.

As stated above, we begin with two, essentially isothermal, spherical, hydrostatic
clusters which merge under the influence of their mutual gravity. The clusters
are composed of two components, one collisional (ICM) and one non-collisional (dark matter).
Although they do interact via gravity, the evolution of these components differ significantly,
particularly during the early stages of the merger.  We begin with the dark matter, or
particle, evolution since it is the most straight-forward. In the higher mass ratio
mergers such as those discussed here, the subcluster is elongated by tidal forces
as it is drawn into primary cluster. When coincident with the primary core, the subcluster is compressed
creating an extremely concentrated mass distribution and very
deep gravitational potential well. Upon
exiting the core, the subcluster is largely disrupted appearing downstream as
a spray of particles some of which are no longer bound to the system.  Those particles
that do remain bound return on radial orbits through the remnant core. The result is
a sustained anisotropy in the particle velocity dispersion that supports an elongation of the
dark matter distribution.  As noted by both Roettiger \etal (1997) and van Haarlem \&
van de Weygaert (1993), the elongation of the dark matter distribution, and, consequently, the
gravitational potential, is long-lived such that the alignment of a cluster merely reflects the
axis of the most recent merger. Ultimately, the ICM will come into equilibrium within
the elongated potential creating a prolate gas distribution. As we will see in \S \ref{results},
this has strong implications for the SZ analysis.

The evolution of the gaseous components is somewhat more complicated. As the subcluster begins
to impinge on the primary cluster, its leading edge is compressed and swept upstream as subcluster gas
is stripped away at large radii. Along the leading edge,  a bow shock is formed as the two cores begin to 
merge which further compresses and heats the gas. At the time of core passage,
the gas distribution is strongly peaked, as is the gravitational potential at this time, and
slightly elongated perpendicular to the merger axis (Figure \ref{imevol}a from Merger 3; Table 1). As the subcluster
exits the core, the initial bow shock proceeds downstream and can be seen as an arc of hot
gas. Behind the shock is a region of rarefaction and adiabatic expansion containing much cooler gas (left edge
of Figure \ref{imevol}d).
As noted above, the gravitational potential reaches a peak minimum during core passage only to
rapidly return to near premerger values as the subcluster exits the core. While near its minimum,
gas is drawn in from the outer regions of the cluster only to be rapidly expelled from the core.
The expelled gas interacts with residual infall from the subcluster creating a second shock
that propagates upstream (far right Figure \ref{imevol}d). 

The fate of the subcluster gas depends on the relative cluster parameters. In general, most
of the subcluster gas is stripped before core passage, remaining as relatively
cool gas on the upstream side of the merger remnant. Gas, along the
leading edge of the subcluster is severely heated (Figure \ref{imevol}b) before being stripped whereupon it is transported 
along the bow shock and deposited in the outer parts of the cluster. In those instances
where the subcluster and primary cluster have comparable ram pressure, some subcluster gas
will penetrate and be carried beyond the primary core. Since the subcluster gas is
no longer bound within its own gravitational potential nor is it  pressure confined by the primary ICM at
large radii, it eventually
expands and cools before falling back into the primary where it is subsequently re-heated and
slowly mixes via random gas motions in the remnant (Figure \ref{imevol}f,h).
All of these effects, the shocks, adiabatic expansion, remnant subcluster gas,  etc, result in
a very inhomogeneous, (\ie non-isothermal) temperature morphology which, as we will see
in \S \ref{results} has implications for the SZ analysis.

\section {RESULTS}
\label{results}

 The origin of the possible discrepancy
in \Hoc at least the component associated with the X-ray observations,
has two primary sources, those being the uncertainty in the true temperature and
density distributions. Here, we attempt to separate  and quantify these two effects
by first presenting an analysis using the ``observed" temperature data, both
a single isothermal temperature and a high-resolution temperature profile
combined with the true LOS density distribution. We then present the
same analysis using the observed temperature data along with an observationally
derived density profile. Since the SZ effect is highly dependent on
the particular LOS through the cluster, we present our results as a function
of projection angle with respect to the observer. And, since the
systems are evolving, we also examine two epochs during
the post-merger evolution. 

\subsection {Temperature}

Figures \ref{stat4b}a-f show the discrepancy between the observed and true value
of \Ho that is due entirely to uncertainty in the LOS temperature distribution as a function
of projection angle. The projection angle is defined relative to 
the merger axis such that +90$^\circ$, 0$^\circ$, and -90$^\circ$ correspond to the
observer looking up the merger axis (subcluster moves toward the observer), perpendicular to, and 
down the merger axis (subcluster moves away from the observer), respectively.
We chose the merger axis as a reference because it is the merger remnant's primary axis of
symmetry. These figures were generated by  calculating the ratio of the observed to the true \Ho (Eq. \ref{ratio})
along 100 random LOS through the cluster core. This results in a distribution of
points to which we fit a parabola. For the sake of clarity, we only plot the parabolic fit.
Typically, the maximum residual to the fit is $\sim\pm$0.10.  The solid line
represents the resulting discrepancy when using
a projected, emission-weighted isothermal temperature estimate taken within a  0.75 Mpc aperture.
Realistically, we can now
expect more detailed temperature information on a significant number of
relatively nearby clusters (Markevitch 1996). To address this possibility, we have performed the SZ analysis using
azimuthally averaged temperature profiles binned by 50 kpc (dashed line).

There are several sources for the scatter in the above relations. Ultimately, they are all tied
to the extreme sensitivity of the denominator of Eq. \ref{ratio} to the particular LOS. As the merger remnant
evolves, the extreme symmetry seen in the early merger stages breaks down giving way to a more
irregular temperature distribution. Consequently, any small variation in the LOS can cause
a significant change in the line integral. On the other hand, the observed temperatures
 are relatively unaffected by projection since they are averaged over large regions of the cluster. 
As stated above, the typical maximum
residual to the fit is $\pm$0.10, this corresponds to $<$10\% variation in the line integral
from one LOS to the next.  Figures \ref{scan}a,b show the LOS variation in $h_{obs}/h_{true}$ for one merger.
In Figure \ref{scan}a, we have chosen LOS in constant latitude ($\perp$ to merger axis) while scanning in
longitude ($\parallel$ to merger axis). In Figure \ref{scan}b, we hold longitude constant and scan in latitude.
Each point represents a LOS.

In order to check for the affects of aperture size on the isothermal temperature, we view
mergers 4 and 5 at one projection (-30$^\circ$, the isothermal temperature is not very strongly
dependent on projection angle) and calculate $h_{obs}/h_{true}$ for various aperture
sizes ranging from 0.25 to 1.25 Mpc. The results of this analysis are shown in Figure \ref{aperture}. We
see a general trend of decreasing $h_{obs}/h_{true}$ with increasing aperture size.
We also note the potential for overestimation of \Ho when using relatively small apertures. Inagaki \etal 1995
observed a similar effect (see \S \ref{comp}). In the presense of a strong temperature gradient,
a small aperture will result in an overestimation of the cluster temperature which
can lead to an overestimation of \Ho by 10\% or more.

Examination of these figures shows 1) \Ho is systematically underestimated in all mergers at all
projection angles. Using an isothermal temperature estimate, we find errors range from $\sim$ 5-30\%.
Using a temperature profile, errors range from $\sim$5-20\%. 2) The systematic underestimation of \Ho tends
to be greater in mergers where the subcluster contains more gas relative to the
primary, Figures \ref{stat4b}a-c,f. 3) Estimates of \Hoc in all but one case, improve
when the temperature profile is used, generally by $\sim$10\%. The improvement is most
noticeable in the more disruptive mergers, (\ie those where the ratio of the central gas densities
is low). Little, if any improvement is noted in the very high gas density ratio mergers, but these
showed relatively small systematic errors to begin with. 4) Generally, the error is relatively flat as a function of projection
angle (latitude), varying by $<$0.1 in $h_{obs}/h_{true}$.

\subsection{Density and Temperature}

Figure \ref{stat4a}a-f shows the discrepancy in \Ho when both the density and temperatures are
observationally determined. As before, the solid line indicates an isothermal temperature
was used, a dashed line indicates a temperature profile. Comparison with Figure \ref{stat4b}a-f reveals the 
relative contribution of each
to the total discrepancy. Several conclusions can be drawn. 1) When viewing along the merger axis ($\pm$90$^\circ$),
the discrepancy increases causing a further underestimation of \Hoc up to 35\% is some cases.
2) When viewing perpendicular to the merger axis (0$^\circ$), the discrepancy is the same as or less than
that due to temperature alone, usually $<$20\%. The combination of (1) and (2) create a signature
by which the \Ho discrepancy is maximum when viewed along the merger axis and at a minimum
when viewed perpendicular to the merger axis. The other trends mentioned above still hold.
In general, the mergers with lower central density ratios have larger systematic errors, and
they benefit most from using a high-resolution temperature profile.

\subsection {Evolution}
We have chosen to perform the SZ analysis at $\sim$2.5 and 5.0 Gyrs 
after core passage. The results can be seen in Figures \ref{hevol}a,b. There were several
reasons for selecting these particular epochs. First, as noted in Roettiger \etal (1997),
much of the internal gas dynamics have diminished by 2.5 Gyrs, particularly bulk flows. 
The core crossing time for these clusters is $\le$1 Gyr which is the absolute lower
limit for post-merger relaxation of the cluster. The actual relaxation time is considerably 
longer owing, at least in part, to continual infalling of subcluster particles and gas for nearly 2 Gyrs.
By waiting until 2.5 Gyrs, we can neglect the kinematic SZ effect (\S \ref{szkin}) in this
portion of the analysis.  A second motivation for the selection of these particular epochs 
is the ill-constrained cluster merger rate.
Both cosmological numerical simulations and observational evidence supports a hierarchical
universe in which largescale structures form and grow via mergers with smaller systems. An
interesting question then pertains to the rate at which these mergers occur. Unfortunately,
ascertaining the true merger rate is very difficult. Based on cooling flow evolution
analysis, Edge \etal (1992) estimate mergers occur
every 2-4 Gyrs. Observational surveys (both optical and
X-ray) show substructure in 30-70\% of clusters (Bird 1993; Mohr \etal 1993; Davis 1994) indicating
that a large fraction of clusters have undergone significant dynamical evolution in
the last 2-3 Gyrs. A theoretical analysis by Richstone, Loeb, \& Turner (1992) shows that
in an $\Omega_o$=1 universe, only 40\% of present day clusters are formed at 80\% of the Hubble time, indicating
significant accretion has occurred within the last 2-4 Gyrs.  In an $\Omega_o$=0.2 universe, about 80\%
of present day clusters are formed by this time. Therefore, we chose 2.5 and 5.0 Gyrs to bracket the time expected between
 significant mergers.

Figure \ref{hevol}a reveals the lack of evolution in Merger 4 (4:1 mass ratio) between 2.5 and 5 Gyrs. As above,
we see the signature of a prolate gas distribution, \ie a severe ($\sim$30\%) underestimation of
\Ho at $\pm90$ projection. This does not appear to change with time indicating
a sustained anisotropy in the particle velocity dispersion  which
supports the elongated mass distribution. This would also seem to indicate very little
evolution in the temperature distribution though only a third of the discrepancy was
attributable to the temperature at $\pm$90$^o$. Figure \ref{hevol}b shows the evolution of the \Ho discrepancy
in Merger 6 (8:1 mass ratio). At 2.5 Gyrs, we see the signature of the prolate gas
distribution. However, at 5.0 Gyrs the signature diminishes noticeably indicating a more
rapid relaxation of the gas distribution within a largely spherical potential.  Once again, there
is little evolution in the temperature distribution, which contributes the bulk of the discrepancy in
this case.

The case for limited evolution is further supported by the values in Table 2. Here, we present the mean
ratio of $h_{obs}/h_{true}$ averaged over all angles at 2.5 and 5.0 Gyrs. In each merger, the
mean ratio is less than unity indicating a systematic underestimation of \Hop However,
there is no clear trend regarding the degree of evolution, except that
mergers involving subclusters with relatively low central gas densities
appear more likely to show significant positive evolution (Mergers 4 and 6 in Table 2). That is,
the observed discrepancy in \Ho decreases with time.

\subsection{Gas Dynamics}
\label{szkin}
In addition to the thermal SZ effect, there is also a kinematic SZ effect which for small values is
an additive term of the form 
 ($\sigma_T/c) \int n_e v_{pec} dl$ where
$v_{pec}$ is the velocity of the scattering plasma relative to the observer
(Sunyaev \& Zeldovich 1980). Therefore,
the estimate of \Ho may be influenced by bulk plasma motions induced along the
observer's LOS during the merger . Roettiger \etal (1997)
showed that mergers can generate bulk flows in excess of 1000 \kms through the cluster
core during the early stages of a merger. This is particularly true in the cases where the
subcluster penetrates the primary before being stripped of its ICM. Observationally, such flows
have not been confirmed, although this situation may change with the launch of the high
resolution X-ray spectrometer, Astro-E.

Unlike the thermal SZ effect, the kinematic effect is directional in that it depends on the direction
of the cluster's motion relative to the observer. Therefore, one would not expect to observe a systematic
effect when studying a large sample of clusters or individual clusters where the gas dynamics
along the LOS are essentially random. However, the kinematic SZ effect can lead to an observational
bias. Clusters  with bulk flows directed away from the observer
will have systematically larger values of $\Delta T/T_{CMB}$ making them easier to detect. This effect is further enhanced by the tendency to select
hotter, more luminous clusters ($L_x \propto T^{5/2}$). Significant mergers will
tend to increase the cluster's
temperature, and consequently increase the X-ray luminosity making them more likely to be
selected in this sample. Furthermore, clusters merging along the LOS will be viewed 
as end-on prolate structures which may be preferentially selected because of their enhanced
X-ray surface brightness.

We address the kinematic SZ effect in merging clusters of galaxies by including an
additive term in the denominator of Eq. \ref{ratio} such that,
\begin{equation}
\label{ratio2}
\frac {h_{obs}} {h_{true}} = \left [ \frac {\left [ 2 \int n_e T_e dl \right ]_{obs} } 
{\left [2 \int n_e T_e dl + \frac{mc}{k} \int n_e v_{pec} dl \right ]_{true} } \right ]^2 ,
\end{equation}
where $v_{pec}$ is the peculiar velocity of a gas parcel along the LOS in the rest frame
of the primary cluster's center of mass. Positive $v_{pec}$
is defined to be motion away from the observer.  We then performed the same analysis as above
on several selected mergers and epochs. In general, we find the kinematic effect is not significant. This
is not surprising, since the bulk flow must not only be high velocity, but coincident with
high density. Only in the early stages of mergers 1 and 2 (those with low ratios of central
gas density) do we find a noticeable effect, 5-6\%,
when observing within 15$^\circ$ of the merger axis and  within 1 Gyr of core passage.

\subsection{A Merger with a Non-Zero Impact Parameter}
\label{offaxis}
We now examine the remnant of a non-zero impact parameter merger. Although cosmological
simulations of largescale structure formation would seem to indicate that most mergers are largely
head-on owing to the infall of matter along radial filaments, the potential
for offaxis mergers does exist. The unusual temperature and density substructure
observed in A754 (Henry and Briel 1995) is believed to be the result of a high
angular momentum merger (Henricksen \& Markevitch 1996). The significance of
offaxis mergers to this study is that residual angular momentum in the remnant may result
in an oblate rather than a prolate gas distribution.

The initial parameters for Merger 7, our offaxis merger, can be seen in Table 1.
Although a somewhat lower mass ratio, Merger 7 is quite similar to Merger 1.
The most significant difference between these two mergers is that in Merger 7
the subcluster is given an initial velocity of 150 \kms perpendicular to the line
of centers. This is in addition to the 300 \kms along the line of centers which was
used in all the merger simulations.
As a result, at the time of closest
approach the respective centers of mass are separated by $\sim$160 kpc or $\sim$0.5$r_c$. 
This is enough to have a significant affect
on the early evolution of the merger remnant.  We find that as the cores interact,
there is a compression and shear which produces a bar of X-ray emission at the interface
between the primary and subcluster gas components. As the subcluster dark matter 
swings around the primary core, it drags ICM with it creating an asymmetric extension in the
X-ray surface brightness distribution. 
Near the leading edge of the extension, there is an arc of shock heated gas. 
Early in the merger evolution, we find that the distinctive X-ray and temperature morphologies 
and the displacement found
between the dark matter and gas distributions are quite similar to 
the observed properties of A754 (Zabludoff \& Zaritsky 1995; Henriksen \& Markevitch 1996). Unlike the
head-on merger, the subcluster is not severely disrupted during the initial passage.
Eventually it reaches turn around and falls back into the core of the primary
on a nearly radial orbit. The second core passage occurs at approximately 2.5
Gyrs after the first. During the first passage angular momentum is transferred
to the gas distribution while some is lost as dark matter particles are shed.
Although the angular momentum in the gas should dissipate quickly there is still some
evidence of rotation at 5.0 Gyrs after the initial core passage. There are rotational
velocities of order 600 \kms at radii of 1.5-3$r_c$ in the plane of the merger. Velocities 
are lower and less ordered within $\sim$1.5$r_c$.

Our SZ error analysis is complicated by the asymmetric evolution of the offaxis merger.
As mentioned above, the second core passage occurs at approximately 2.5 Gyrs after
closest approach. Consequently, the cluster morphology is still significantly 
disrupted at this time, and it is certainly not recommended that such a cluster be included
in an SZ-\Ho study. Still, for the sake of completeness and consistency, we perform the above analysis
at 2.5 Gyrs. We find that the results are quite consistent with those of the head-on
merger. This is as expected since the remnant is still largely prolate. There is, however,
some oblateness which causes an overestimation of \Ho by as much as 20\% when the cluster is
viewed  within 15-30$^\circ$ of face-on (\ie perpendicular to the oblateness).  This is
caused by an overestimation of the LOS pressure integral. Overestimations
of both temperature and density contribute to this effect. The gas distribution that
is projected into the plane of the sky appears to be
more extended than it actually is along the observer's LOS, and the temperature is less 
contaminated by
cooler gas along the LOS and therefore appears somewhat higher.  The analysis was also performed 
at 5.0 Gyrs after closest approach. Here, the results are even more similar to
those of Merger 1 although the residual oblateness does reduce the underestimation of \Ho
by 5-10\% within about 20$^\circ$ of face-on. This does not however significantly affect the
\Ho discrepancy when averaged over all viewing angles, compare Merger 7 at 5 Gyrs with Merger 1
at 5 Gyrs in Table 2.

\subsection{A Comparison with a Previous Study}
\label{comp}

Inagaki \etal (1995) also addresses  the reliability of \Ho derived from the SZ effect.
Our study differs from theirs in several significant ways. First, our numerical methodology 
for computing the evolution of clusters (\S \ref{num})
is based on PPM where as they employ the numerical simulations of Suginohara (1994) which
are based on a smoothed particle hydrodynamics (SPH) code. PPM is better able to
accurately record the development and evolution of shocks which can dramatically effect the
temperature evolution within merging clusters of galaxies. Second, we have effectively twice
the resolution (50 kpc) over a much larger region (2 Mpc wide, extending the length of our
grid, \S \ref{num}). Inagaki \etal report 100 kpc resolution, but this really only applies to 
the core of the cluster. As density decreases, so does their resolution.  Third, our approach to
the analysis is considerably different. Inagaki \etal observed two simulated clusters evolved
from cosmological initial conditions but without noting their recent history. We have chosen
to examine a series of clusters with known initial properties (based on the observed
properties of clusters) and very specific merger histories.
In this way, we can directly relate the observed \Ho discrepancy to the cluster's recent evolution.
Finally, we have adopted a more systematic approach to presenting the analysis in order
to elucidate projection effects under specific observational conditions. The Inagaki \etal
study was directed toward a more statistical analysis which would be applicable to a large
sample of SZ-based \Ho observations. Our study is meant to show the magnitude of errors
attributable to any given observation.

With this said, we feel our results are quite consistent with, although somewhat more
extreme than, those of Inagaki \etal.
They report that non-isothermality of clusters, particularly temperature gradients, is 
the most significant source of error in \Hoc resulting in as much as a 20\% 
 underestimation. We find up to 30\% in our merger simulations though $<$20\% was
more typical. They also note that prolateness of the cluster will cause an underestimation
of \Ho when viewing the cluster along the major axis.  However, because of the formalism they
choose for {\it calculating} the core radius, they attribute the underestimation of \Ho to an overestimation 
of the core radius. We, on the other hand, find that {\it fitting} a core radius and $\beta$ to
the X-ray surface brightness of a prolate gas distribution viewed end-on effectively results in 
an underestimation of the LOS extent of the gas distribution.
Combined with the temperature-based
underestimation, we find a combined effect of up to 35\%. They quantify the error due
to clumpiness and asphericity as a $<$15\% overestimation. We also see the potential for overestimation 
of \Ho due to both asphericity and clumpiness. We find viewing a prolate distribution,
perpendicular to the major axis can result in a overestimation of 10-15\%. In some instances,
this is just enough to compensate for the underestimation of \Ho due to incomplete temperature
information.

\section {DISCUSSION}
\label{discus}

\subsection{Origin of the \Ho Discrepancy}

The origin of the \Ho discrepancy in these dynamically evolving clusters
can be summarized simply as a lack of true 3-dimensional information and the
volume-weighted observed quantities. The LOS projection
effects and limited resolution of the X-ray observations 
tend to mitigate extremes of temperature. As seen from Eq. \ref{dt},
the SZ effect is most influenced by the hottest, densest gas in the cluster. And, although
the observed temperature is emission-weighted, the emissivity is only weakly temperature
dependent. This, combined with the long LOS through the cluster, causes the observed temperatures
to be lower than the effective mean temperature of the scattering plasma.
In the presense of even a mild temperature gradient, an isothermal temperature estimate
using a large aperture will include a large amount of gas at relatively cool temperatures
compared to the core, thus weighting the mean temperature downward.  Underestimation of
the temperature, according to Eq. \ref{heqn}, will lead to an underestimation
of \Hop  As the aperture size increases, the observed temperature decreases which further reduces the
observed value of \Ho (see Figure \ref{aperture}). Lack of spherical symmetry,
and the extremely local nature of inhomogeneities in the temperature distribution cause
even the temperature profile to differ significantly from the true LOS temperature experienced
by the CMB photons.  This is best demonstrated in Figures \ref{slice1}a and \ref{slice2}a where we show the observed isothermal
cluster temperature within 0.75 Mpc, the azimuthally averaged temperature profile, and the true
temperature along the observed LOS both perpendicular and parallel to the merger axis from Merger
4 at 5 Gyrs.  Note that the 
temperature profile
appears largely isothermal varying by little more than 1 keV across the face of the cluster.
In stark contrast, the  true LOS temperature exhibits several local temperature peaks.

Similarly, projection effects play an important role in the density analysis. Each
of the mergers discussed here results in a prolate gas distribution whose
major axis coincides with the merger axis. Both the degree
of ellipticity and its evolution are a function of the relative total mass and the relative
central gas densities. As the subcluster increases both in gas content and total mass
relative to the primary, the degree of prolateness in the gas distribution increases.
Initially, the gas distribution is more elongated than the dark matter distribution.
As the systems come into equilibrium, the gas traces the dark matter fairly well.
In the case of the 4:1 mass ratio mergers both the dark matter and gas distribution
remain noticeably elongated even after 5 Gyrs.  In the 8:1 mass ratio mergers, both
the gas and dark matter relax into essentially spherical distributions.

The signature of the prolate gas distribution is easily seen in the results of the
SZ analysis. When viewing the elongated cluster along the major axis, projection effects
cause the gas distribution to appear less extended and more centrally concentrated than
in reality. Consequently, both core radius and $\beta$  fit to the X-ray profile are too small. 
This causes the central gas density in Eq. \ref{sx} to be slightly overestimated, but
the primary effect is to underestimate the extent of the gas distribution and ultimately \Hop The effect
is reversed when viewing along the cluster's minor axis. Here, the tendency is to overestimate $r_c$ and
$\beta$ causing a slight underestimation of the central gas density, but this effect is 
more than compensated for by the overestimation of the extent of the gas distribution along
the LOS. This leads to an overestimation of \Hoc but it is usually not enough to compensate for
the underestimate resulting from uncertainties in the temperature distribution.  Figures \ref{slice1}b
and \ref{slice2}b
show a comparison of the true LOS density profile (solid line) and the observationally
derived profiles parallel and perpendicular to the merger axis, respectively. This effect varies fairly
smoothly between these two extremes resulting in the distributions plotted in Figures \ref{stat4a}a-f.

\subsection{Limitations of this Analysis}
There are several potentially important physical processes neglected in this study
which may affect the determination of \Ho by the SZ effect.
For example, we do not include radiative cooling or thermal conduction, both of which may
serve to mitigate some of the effects observed here. (Similarly, Inagaki \etal (1995)
also neglect these processes).  This is particularly true of
thermal conduction. As with radiative cooling, thermal conduction is likely to be most
important in the dense cores of clusters, precisely the region that most influences the
X-ray emission while also strongly influencing the SZ effect. If thermal conduction in clusters is characterized by the Spitzer (1962)
coefficient of thermal conductivity, temperature inhomogeneities will likely be erased
on relatively short time scales, $<$1 Gyr (Sarazin 1986). However, conduction may be very much limited
to the core owing to the strong dependence on density and therefore radius. Still, the actual
level of thermal conductivity, which may depend on the detailed structure and strength 
of intracluster magnetic fields (Tribble 1989), is uncertain by at least two orders of magnitude. For this
reason, it is very difficult to address conduction with numerical simulations at this time
simply because the parameter space is so poorly constrained.
Radiative cooling in the ICM is better understood, but it is difficult to gauge the affect
on this particular analysis.  Cooling may serve to weaken radial temperature
gradients, and it may also result in small scale clumping which poses other problems
for the SZ analysis. We have chosen our parameter space such that cooling
is not significant. Cooling times range from $\sim$5 Gyrs to greater than a Hubble time.

\section {Summary}
\label{conclusions}
We have shown that significant systematic errors in the SZ-based value of \Ho can result
from a combination of non-isothermality and asphericity in both the temperature and
density distributions resulting from recent dynamical evolution. Together, these factors
can cause \Ho to be underestimated by as much as 35\% with more typical values ranging from
10-25\%. Although there is a potential for overestimation of \Ho under specific conditions
(in particular, oblate clusters resulting from offaxis mergers when viewed nearly face-on),
the tendency to underestimate is far more prevalent.  This effect will be enhanced by various
selection effects which may bias SZ samples toward clusters with larger (\ie more detectable)
$\Delta T/T_{CMB}$. For a given set of X-ray properties, the observed \Ho estimate
will decrease as $\Delta T/T_{CMB}$ increases. Similarly, X-ray based selection effects
will also tend to bias toward an underestimation of \Hop Merger remnants tend to 
be hotter and more luminous which increases their likelihood of being selected. Remnants of head-on
or nearly head-on mergers will be 
 prolate. Prolate clusters viewed end-on will be preferentially selected by
virtue of an enhanced X-ray surface brightness.

We find that detailed temperature profiles do not significantly reduce the error in
\Hoc except in the more extreme merger examples. It is important to note however
that no X-ray telescope will supply the true 3-dimensional temperature (or density) information, so
a reliance on spherical symmetry will remain important. In order to minimize the effects
of dynamical evolution, a statistical sample of clusters having SZ measurements
should be compiled. The members of this sample should appear spherically symmetric in
both their X-ray surface brightness, and temperature distribution. Elongated clusters with
twisted isophotes and multiple emission peaks should be avoided. Also clusters with
large $\beta$ discrepancies ($\beta_{fit} \ne \beta_{spec}$, Sarazin, 1986) as well as those with
$\beta_{spec}$ much less than or much greater than unity
should be avoided. Anisotropy in the galaxy velocity
distribution can be the signature of merger activity and when projected into the plane
of the sky can result in a large $\beta$ discrepancy (\eg A2256; Roettiger \etal 1995).
It will be beneficial for future studies to focus on nearby cluster samples which
are less subject to observational selection effects while having more detailed X-ray data.
Finally, we suggest that the numerical modeling of specific systems, such as was done with
A2256 (Roettiger \etal 1995),  will aid in the interpretation of the SZ base determinations
of \Hop

\begin{center}{\it Acknowledgements}
\end{center}
This work was funded by NASA GSFC through RFM as XMM mission scientist.  
We thank the Earth and Space Data  Division of NASA-GSFC for use of the MasPar MP-2.  We thank D. Spergel and
T. Herbig for their comments on the manuscript.
We also thank K. Olson for contributing the particle-mesh code.
Finally, we thank the anonymous referee for a careful reading
of the manuscript.
\newpage
\onecolumn
\begin{table}
\label{tab1}
\begin{center}
\begin{tabular}{c c c c c c c c c } 
\multicolumn{9}{c}{Table 1.  Cluster Parameters}\\ \hline \hline
\multicolumn{1}{c}{Merger} &
 \multicolumn{1}{c}{Cluster} &
 \multicolumn{1}{c}{M(R$<$3 Mpc)} &
 \multicolumn{1}{c}{$n_{e}(R=0)$} &
 \multicolumn{1}{c}{$T_e$ (R=0)} &
 \multicolumn{1}{c}{$r_c^b$} &
 \multicolumn{1}{c}{$f_g^c$} &
 \multicolumn{1}{c}{$\beta_{spec}^d$}&
 \multicolumn{1}{c}{ $n_{e,P}/n_{e,S}^e$}\\
\multicolumn{1}{c}{ ID } &
 \multicolumn{1}{c}{ ID$^a$ } & 
 \multicolumn{1}{c}{10$^{14}$ M$_\odot$} &
 \multicolumn{1}{c}{ $10^{-3}$ cm$^{-3}$ } &
 \multicolumn{1}{c}{ keV } &
 \multicolumn{1}{c}{ kpc } &
 \multicolumn{1}{c}{ R $<$ 1 Mpc } &
 \multicolumn{1}{c}{   }&
 \multicolumn{1}{c}{ Ratio  }  \\ \hline

1 & P  &  15.0 & 5.0 &  9.3  & 300 & 0.12 & 0.7 & 1  \\
  & S  &  3.75 & 5.0 &  3.7  & 190 & 0.12 & 0.7 &   \\
2 & P  &  15.0 & 5.0 &  9.3 & 300 & 0.12 & 0.7 & 1.7 \\
  & S  &  3.75 & 3.0 &  3.7 & 190 & 0.09 & 0.7 &  \\
3 & P  &  15.0 & 9.7 &  6.5 & 300 & 0.12 & 1.0 & 2.7 \\
  & S  &  3.75 & 3.7 &  3.4 & 190 & 0.09 & 0.75&   \\ 
4 & P  &  15.0 & 18.0 & 6.5 & 300 & 0.22 & 1.0 & 4.9  \\
  & S  &  3.75 & 3.7 &  3.4 & 190 & 0.09 & 0.75 &    \\ 
5 & P  &  15.0 & 9.7 &  6.5 & 300 & 0.13 & 1.0 &  2.6 \\
  & S  &  1.87 & 3.8 &  2.0 & 165 & 0.15 & 0.75 &     \\ 
6 & P  &  15.0 & 9.7 &  6.5 & 300 & 0.13 & 1.0 & 11.8 \\
  & S  &  1.87 & 0.8 &  2.9 & 165 & 0.08 & 0.5 &   \\ 
7 & P  &  15.0 & 5.0 &  9.3 & 300 & 0.12 & 0.75 & 1  \\
  & S  &   6.0 & 5.0 &  5.0 & 200 & 0.08 & 0.78 &    \\ \hline 
\end{tabular}
\end{center}
\caption[Sample Parameters]
{$^a$ P=primary S=subcluster.
$^b$ Core radius. 
$^c$ gas fraction by mass within 1 Mpc.
$^d$ $\beta_{spec} = \mu m_p {\sigma_v}^2/ k T)$.
$^e$ Ratio of central electron densities Primary/Subcluster.}
\end{table}

\newpage
\begin{table}
\label{tab2}
\begin{center}
\begin{tabular}{c c c c } 
\multicolumn{4}{c}{Table 2.  $<h_{obs}/h_{true}>_\Omega$}\\ \hline \hline
\multicolumn{1}{c}{Merger} &
 \multicolumn{1}{c}{Time$^a$ (Gyrs)} &
 \multicolumn{1}{c}{Isothermal T$^{b,d}$} &
 \multicolumn{1}{c}{T(r)$^{c,d}$}\\
1 & 2.5 & 0.72$\pm$0.05 & 0.82$\pm$0.06 \\
  & 5.0 & 0.64$\pm$0.05 & 0.77$\pm$0.04   \\
2 & 2.5 & 0.66$\pm$0.04 & 0.83$\pm$0.05  \\
  & 5.0 & 0.69$\pm$0.05 & 0.81$\pm$0.05     \\
3 & 2.5 & 0.81$\pm$0.11 & 0.90$\pm$0.09  \\
  & 5.0 & 0.72$\pm$0.09 & 0.90$\pm$0.09     \\
4 & 2.5 & 0.87$\pm$0.10 & 0.88$\pm$0.10  \\
  & 5.0 & 0.88$\pm$0.10 & 0.92$\pm$0.08   \\
5 & 2.5 & 0.80$\pm$0.07 & 0.95$\pm$0.07  \\
  & 5.0 & 0.72$\pm$0.05 & 0.82$\pm$0.06  \\
6 & 2.5 & 0.87$\pm$0.06 & 0.80$\pm$0.07  \\ 
  & 5.0 & 0.90$\pm$0.06 & 0.95$\pm$0.05   \\ 
7 & 5.0 & 0.65$\pm$0.08 & 0.79$\pm$0.09  \\ \hline
\end{tabular}
\end{center}
\caption[Sample Parameters]
{The mean ratio of the observed to true \Ho for each of the first 6
 mergers in Table 1, averaged over all projection
angles at 2.5 and 5.0 Gyrs after core passage. Merger 7 was only analyzed at 5.0 Gyrs.
$^a$ Time since core passage in Gyrs.
$^b$ Ratio computed using an isothermal temperature
estimate. $^c$ Ratio computed using a high resolution
temperature profile. $^d$ Uses observed
density profile based on an isothermal $\beta$ model.}
\end{table}

\newpage
\begin{center}{REFERENCES}
\end{center}
 \everypar=
   {\hangafter=1 \hangindent=.5in}

{Bahcall, N.  \& Cen, R. 1993, ApJ, 407, L49

Bird, C. M. 1993, Ph. D. Thesis, University of Minnesota, Minneapolis

Birkinshaw, M., Gull, S. F., \& Hardebeck, H. 1984, Nature, 309, 34

Birkinshaw, M., Hughes, J. P., Arnaud, K. A. 1991, ApJ, 379, 466

Birkenshaw, M. \& Hughes, J. P. 1994, ApJ, 420, 33

Briel, U. G. \& Henry, J. P. 1994, Nature, 372, 439

Carlstrom, J. E., Joy, M., \& Grego, L. 1996, ApJ, 456, L75

Colella, P. \& Woodward, P. 1984, J. Comp. Phys. 54, 174.

Davis, D. 1994, Ph.D. Thesis, University of Maryland, College Park

Doe, S., Ledlow, M., Burns, J. O., \& White, R. 1995, AJ, 110, 46

Edge, A. C., Stewart, C. G., \& Fabian, A. C. 1992, MNRAS, 252, 414

Freedman \etal 1994, ApJ, 427, 628

Hamuy, M., Phillips, M. M., Maza, J., Suntzeff, N. B., Schommer, R. A., 
\& Aviles, R. 1995, AJ 109, 1

Henriksen, M. J. \& Markevitch, M. 1996, ApJ, 443, L79

Henry, J. P., \& Briel, U. G. 1995, ApJ, 443, L9

Henry, J. P., \& Henriksen, M. J. 1986, ApJ, 301, 689

Herbig, T. 1996, private communication.

Herbig, T., Lawrence, C. R., Readhead, A. C. S., \& Gulkis. S. 1995,
ApJ, 449, L5

Honda, H., Hirayama, M., Watanabe, M., Kunieda, H., Tawara, Y., 
Yamashita, K., Ohoshi, T., Hughes, J. P., \& Henry, J. P., 1996 preprint

Hughes, J. P. 1989, ApJ, 337, 21

Inagaki, Y., Suginohara, T. \&  Suto, Y. 1995, PASJ, 47, 411

Jones, C., \& Forman, W. 1991, BAAS, 23, 1338

Jones, M., Saunders, R., Alexander, P., Birkinshaw, M., Dillon, N., Grainge, K.,
Hancock S., \& Lasenby, A. 1993, Nature, 365, 320

Kobayashi, S. Sasaki, S. \& Suto, Y. 1996, PASJ, in press

Lubin, L. M., Cen, R., Bahcall, N. A., \& Ostriker, J. P. 1996, ApJ, 460, 10

Markevitch, M. 1996, ApJ, 465, L1

Mohr, J. J., Fabricant, D. G., \& Geller, M. J. 1993, ApJ, 413, 492

Mohr, J. J., Evrard, A. E., Fabricant, D. G., \& Geller, M. J. 1995, ApJ, 447, 8

Mould, J. \etal 1995, ApJ, 449, 413

Mulchaey, J. S., Davis, D. S., Mushotzky, R. F., \& Burstein, D. 1996, ApJ,
456, 80

Mushotzky, R. F., Lowenstein, M., Arnaud, K.A., Tamura, T., Fukazawa, Y.,
 Matsushita, K., Kikuchi, K., \& Hatsukade, I. 1996, ApJ, 446, 686

Phillips, M. 1993, ApJ, 409, 14  

Pierce, M. \& Tully, R. B. 1988, ApJ, 330, 588

Pierce, M., Welch, D., van den Bergh, S., McClure, R. Racine, R., \& Stetson,
P. 1994, Nature, 371, 385

Rephaeli, Y. 1995, ARAA, 33, 541

Richstone, D., Loeb, A. \& Turner, E. L. 1992, ApJ, 393, 477

Riess, A., Press, W., \& Kirshner, R. 1995, ApJ, 438, L17

Roettiger, K., Burns, J. O., \& Loken, C. 1993, ApJ,  407, L53

Roettiger, K. Burns, J. O., \& Pinkney, J. 1995, ApJ, 453, 634

Roettiger, K., Burns, J. O., \& Loken, C. 1996, ApJ, 473, 651 

Roettiger, K., Loken, C., \& Burns, J. O. 1997, ApJS, in press

Sarazin, C. 1986, Rev. Mod. Phys., 58, 1

Schaeffer, B. E. 1996, ApJ, 460, L19

Schmidt, B. P., Kirshner, R. P., Eastman, R. G., Phillips, M. M., Suntzeff, N. B.
Hamuy, M., Maza, J., \& Aviles, R. 1994, ApJ, 432, 42

Spitzer, L 1962, {\it Physics of Fully Ionized Gases}, (New York: Wiley)

Suginohara, T. 1994, PASJ, 46, 441

Sunyaev, R. A. \& Zeldovich, Y. B. 1972, Comm. Astrphys. Sp. Phys., 4, 173

Sunyaev, R. A. \& Zeldovich Y. B. 1980, MNRAS, 190, 413

Tonry, J. 1991, ApJ, 373, L1

Tribble, P. 1989, MNRAS, 238, 1247

van Haarlem, M. \& van de Weygaert, R. 1993, ApJ, 418, 544

Watt, M. P., Ponman, T. J., Bertram, D., Eyles, C. J., Skinner, G. K., \&
Willmore, A. P. 1992, MNRAS, 258, 738

White, D. A. \& Fabian, A. C. 1995, MNRAS, 273, 72

Woodward, P. \& Colella, P. 1984, 54, 115

Zabludoff, A. I. \& Zaritsky, D. 1995, ApJ, 447, L21}

\newpage

\begin{center}{\bf FIGURE CAPTIONS}
\end{center}

\begin{figure}[htbp]
\centering \leavevmode
\epsfxsize=0.7\textwidth \epsfbox{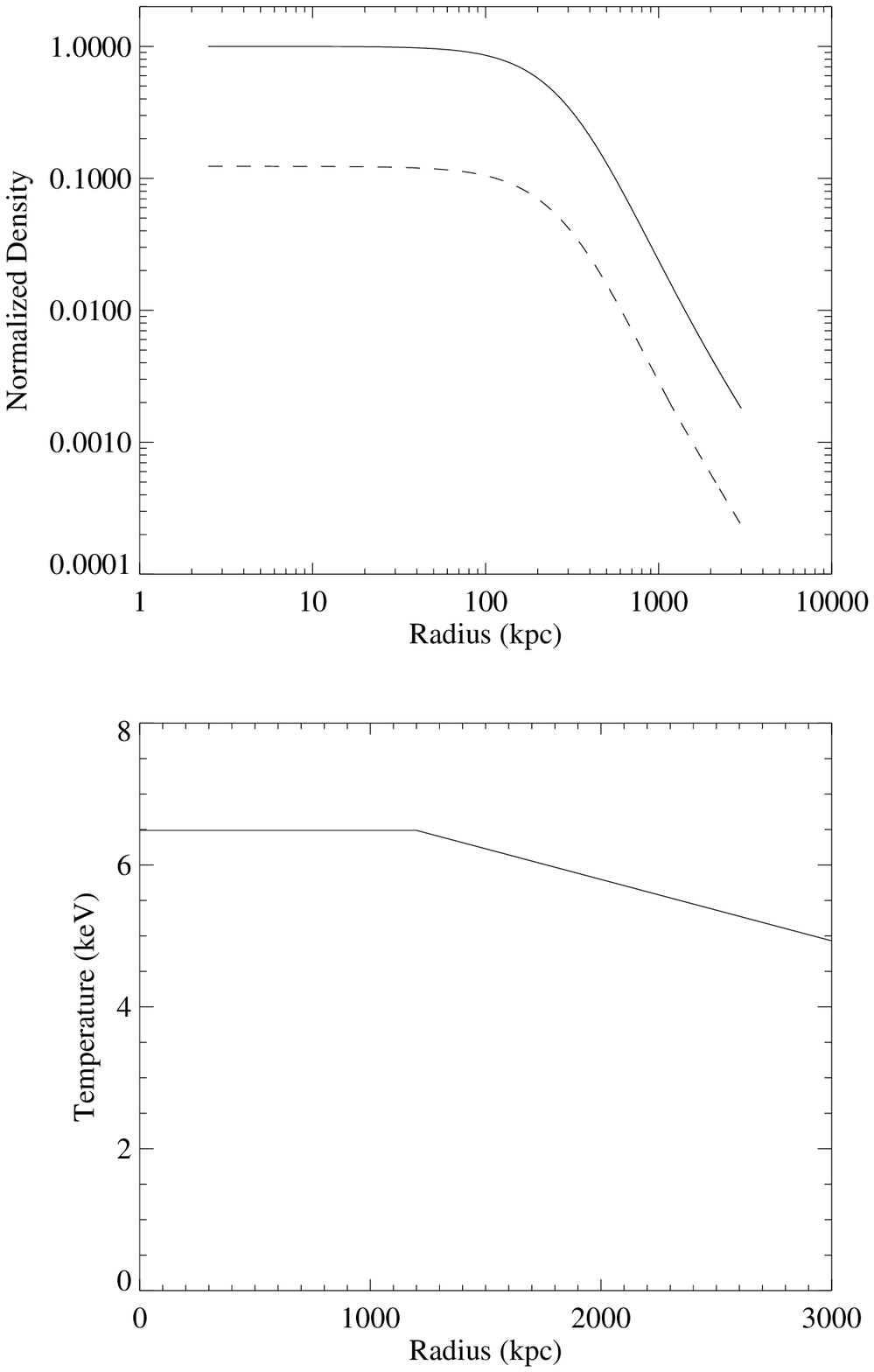}
\caption[]
{ }
\label{initprof}
\end{figure}

{\bf Fig. \ref{initprof}:} a) The normalized initial total mass density (solid)
and gas density (dashed) for the primary cluster in mergers 2-6 (See Table 1).
The shape of the total density profile is identical for all simulated clusters, both
primary and subclusters, only the dimensions and total mass scaling change.
The gas density profiles change only slightly. This cluster has $\beta$=1. Those
with lower $\beta$ values have flatter gas density profiles for $r>r_c$.
b) The temperature profile for the same cluster as in (a). The cluster
is isothermal for $r<4r_c$ with a slight temperature gradient at larger radii.
The other clusters have the same basic temperature profile although scaled
in accordance with the choice of $\beta$ and the total cluster mass.

\begin{figure}[htbp]
\centering \leavevmode
\epsfxsize=0.7\textwidth \epsfbox{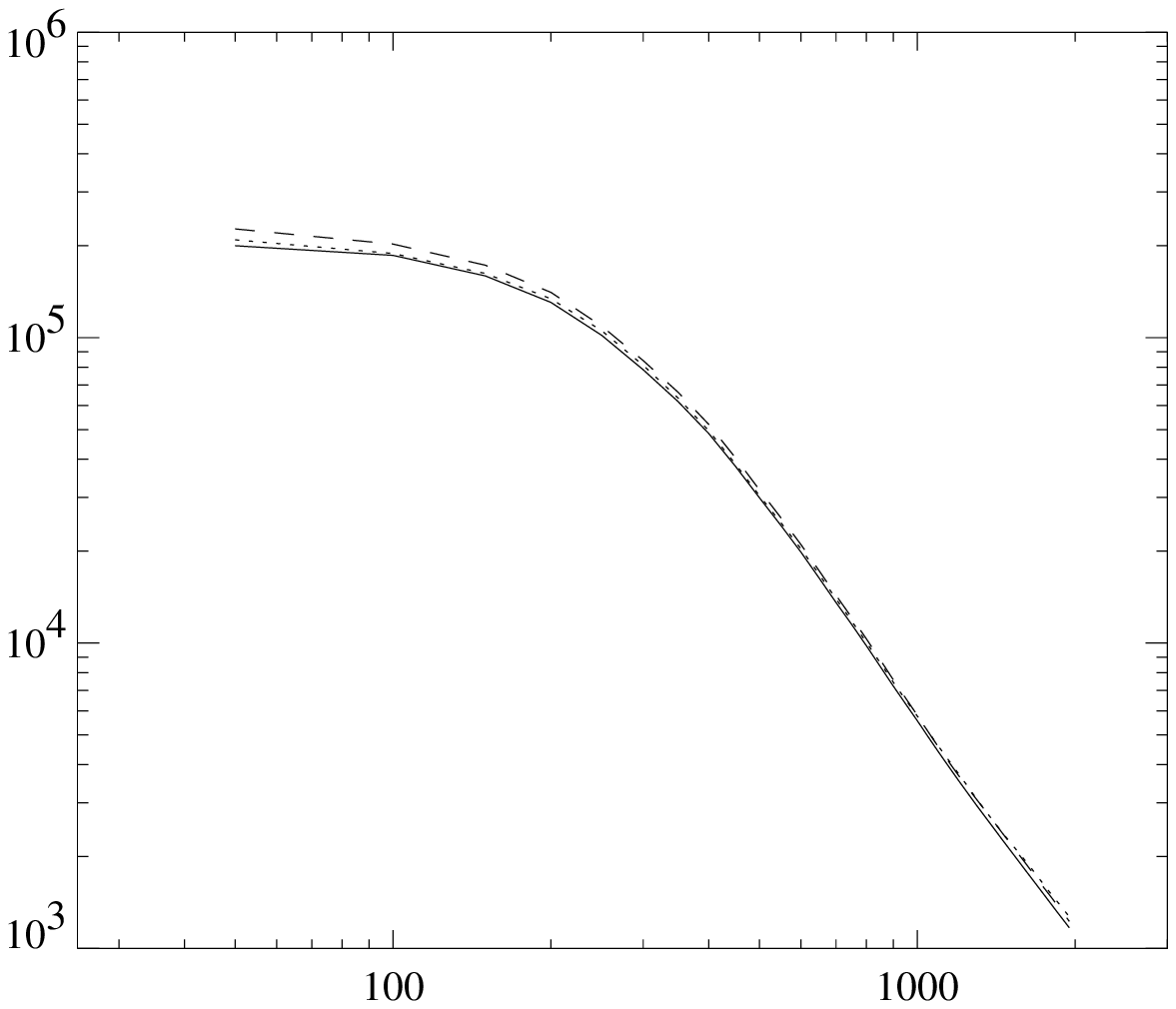}
\caption[]
{ }
\label{sngled}
\end{figure}

\bigskip
{\bf Fig. \ref{sngled}:} Evolution of the isolated cluster's gas density profile.
At 0.0 Gyrs (solid), at 4.0 Gyrs (dotted), and at 8 Gyrs (dashed). Core passage for
the mergers occurs at $\sim$4.0 Gyrs.

\begin{figure}[htbp]
\centering \leavevmode
\epsfxsize=0.9\textwidth \epsfbox{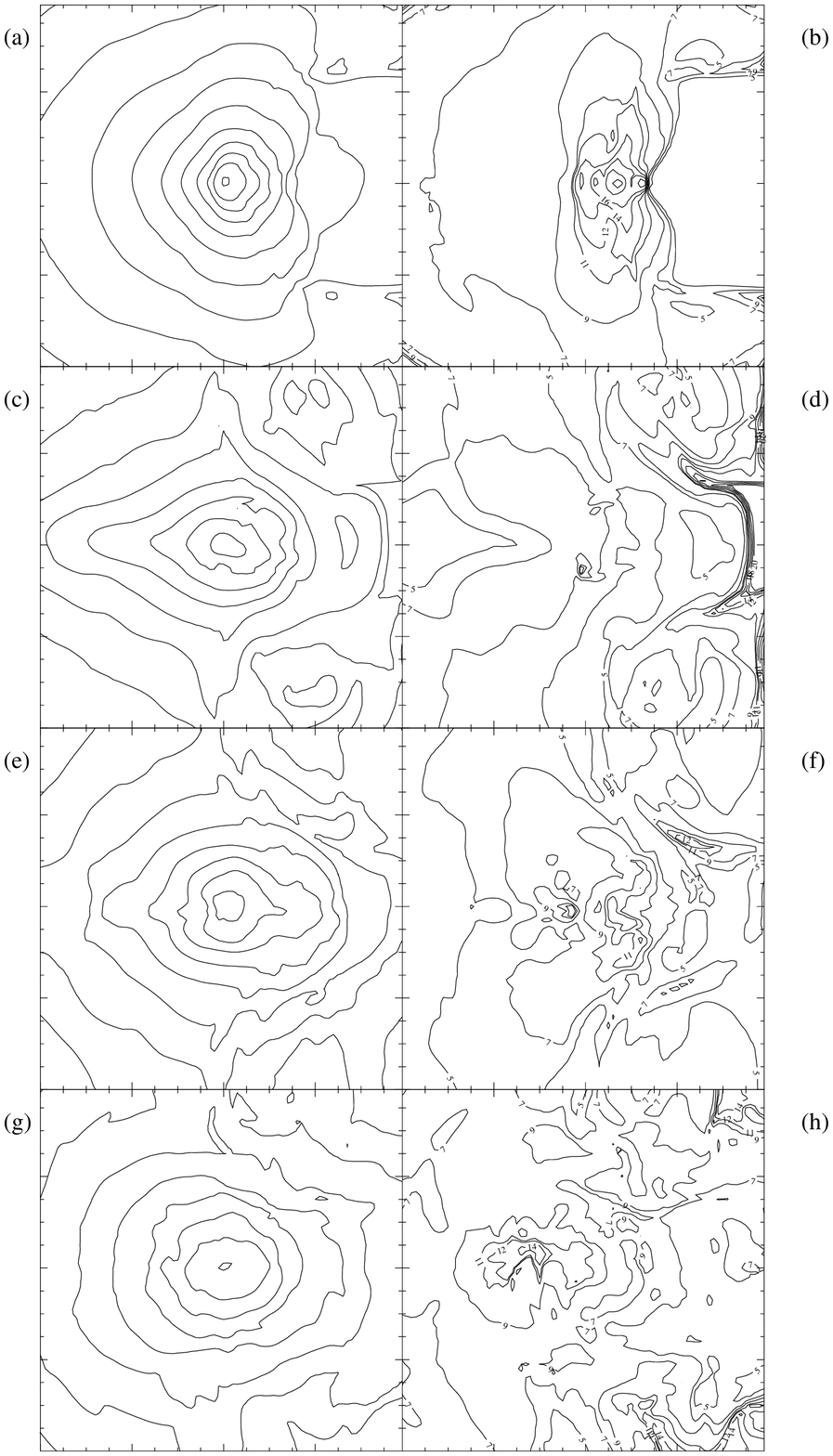}
\caption[]
{ }
\label{imevol}
\end{figure}

\bigskip
{\bf Fig. \ref{imevol}:} Contours of a 2-dimensional slice in gas density (Merger 3, Table 1)
taken through the cluster core and parallel to the merger axis at a) 0.0
Gyrs, c) 1.25 Gyrs, e) 2.5 Gyrs, and g) 5.0 Gyrs. Times are relative to
core passage. Contours are uniformly
spaced in the logarithm and span 2 orders of magnitude. The subcluster
entered from the right moving to the left. The corresponding
slices in gas temperature can be seen in (b),(d),(f), and (h). Contours are linear
and most are labeled with the corresponding temperature in keV. All contours are
spaced by 2 keV.   Each panel
is 4 Mpc on a side.

\begin{figure}[htbp]
\centering \leavevmode
\epsfxsize=0.8\textwidth \epsfbox{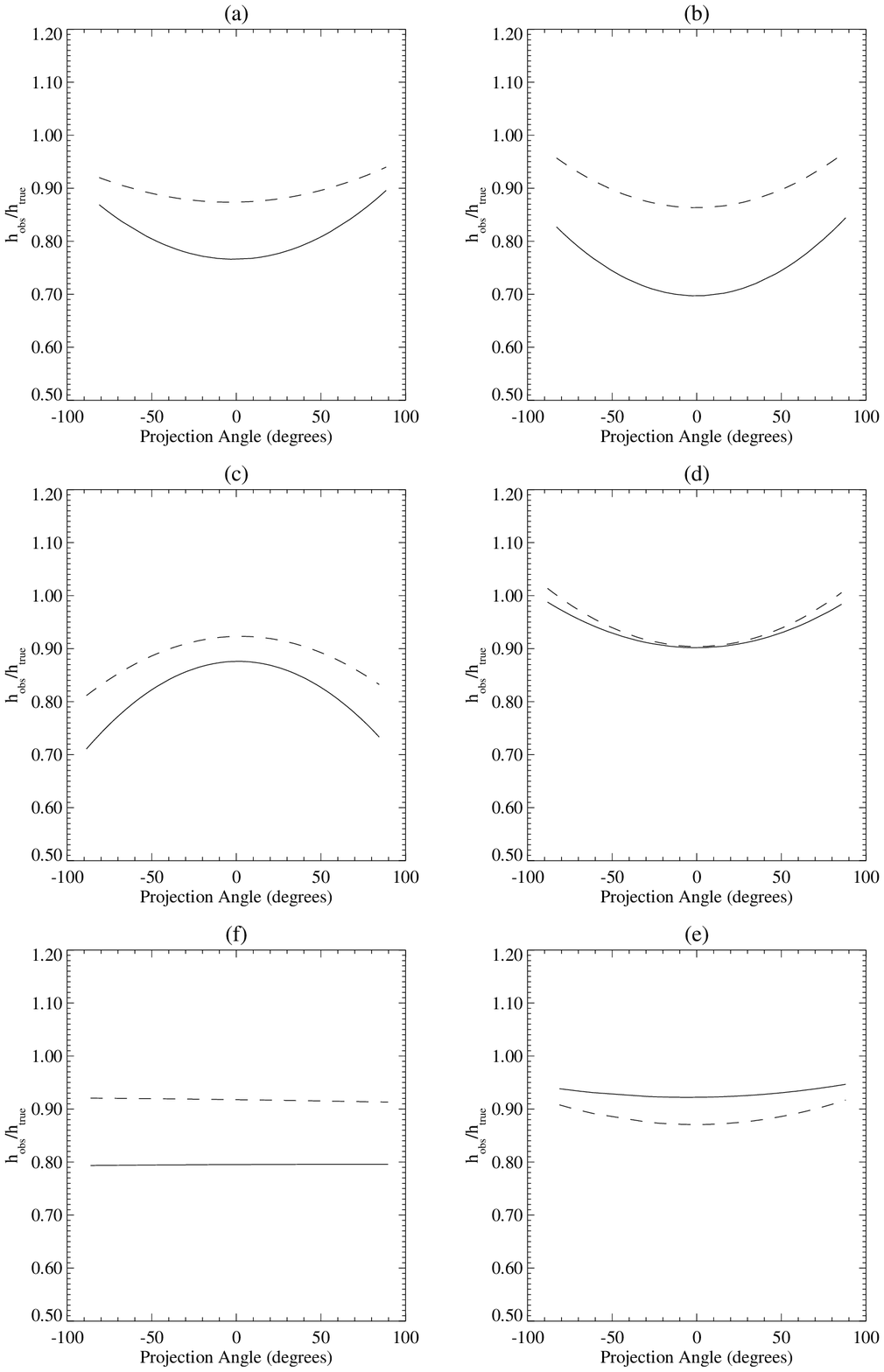}
\caption[]
{ }
\label{stat4b}
\end{figure}

\bigskip
{\bf Fig. \ref{stat4b}:} The ratio of the ``observed" to ``true" \Ho defined by
Eq. \ref{ratio} using the emission-weighted isothermal temperature (solid line)
and the projected, emission-weighted temperature profile (dashed line) as a function of
projection angle along the merger axis (also cluster's major axis). Lines
represent a fit to the distribution of randomly drawn LOS through the cluster
core. Maximum residuals to the fit are $\sim$0.1. In both cases, the exact $n_e$ distribution is used.
A projection angle of 0$^\circ$ implies the observer is looking perpendicular
to the merger axis, +90$^\circ$ is looking up the merger axis (subcluster moving toward
the observer), -90$^\circ$ is looking down the merger axis (subcluster moving away from
the observer). Each panel a-f
corresponds to a merger, 1-6, in Table 1.

\begin{figure}[htbp]
\centering \leavevmode
\epsfxsize=0.7\textwidth \epsfbox{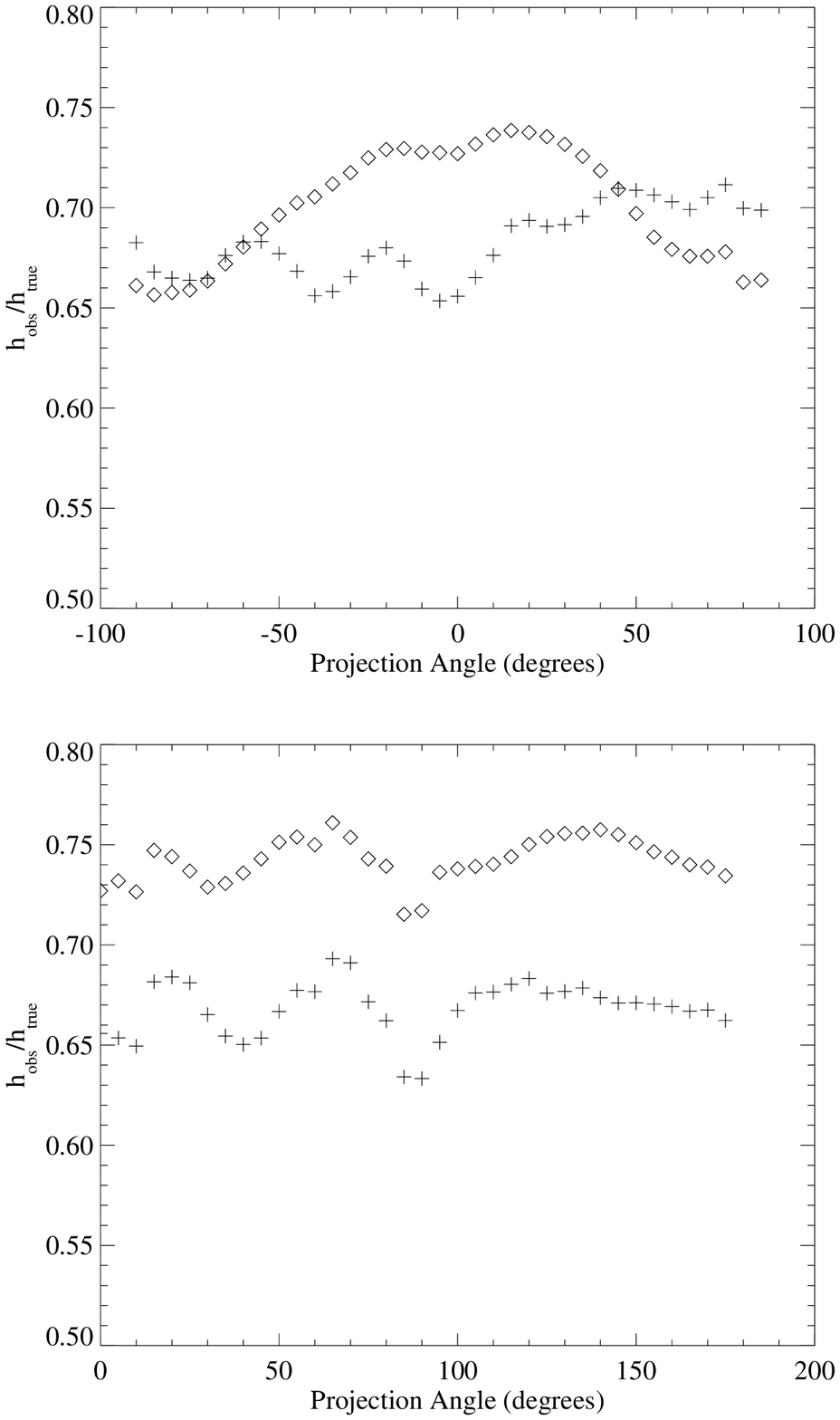}
\caption[]
{ }
\label{scan}
\end{figure}

\bigskip
{\bf Fig. \ref{scan}:} The ratio of the ``observed" to ``true" value of \Ho
defined by Eq. \ref{ratio} using observationally derived temperature {\it and} density 
information as a function of
projection angle  (a) along the merger axis (\ie latitude) at fixed longitude (0$^\circ$) and (b)
perpendicular to the merger axis (\ie longitude) at fixed latitude (0$^\circ$) (Merger 5, Table 1). 
Each point represents a single 
measurement. The + designates use of an isothermal temperature. The $\diamond$ designates
use of an azimuthally averaged temperature profile. The projection angle in (a) is
defined as it is in Figure \ref{stat4b}. The zero point for the projection angle
in (b) is arbitrary owing to the relative symmetry about the merger axis.  Variation of
$h_{obs}/h_{true}$ from one pointing to the next is a result of local inhomogeneities
in the temperature and density distributions.

\begin{figure}[htbp]
\centering \leavevmode
\epsfxsize=0.9\textwidth \epsfbox{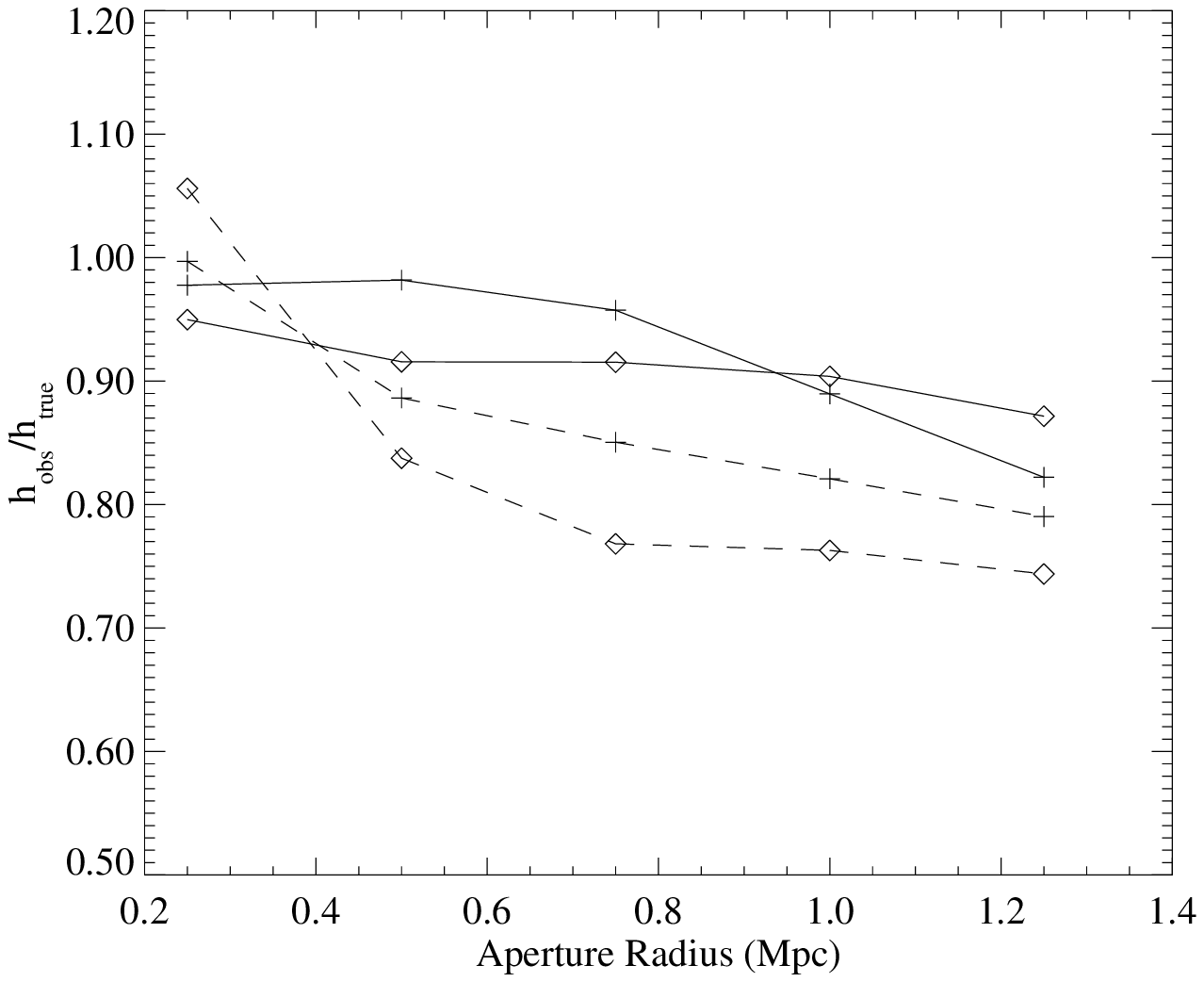}
\caption[]
{ }
\label{aperture}
\end{figure}

\bigskip
{\bf Fig. \ref{aperture}:} The ``observed" discrepancy in \Ho resulting
from limited temperature information only as a function of aperture size
for mergers, 4 (+) and 5 ($\diamond$) at 2.5 Gyrs (solid line) and 5 Gyrs (dashed line).  As the aperture increases,
temperature gradients generated during the merger, cause the observed temperature
to decrease. As the temperature decreases so does the observed value of \Ho.

\begin{figure}[htbp]
\centering \leavevmode
\epsfxsize=0.8\textwidth \epsfbox{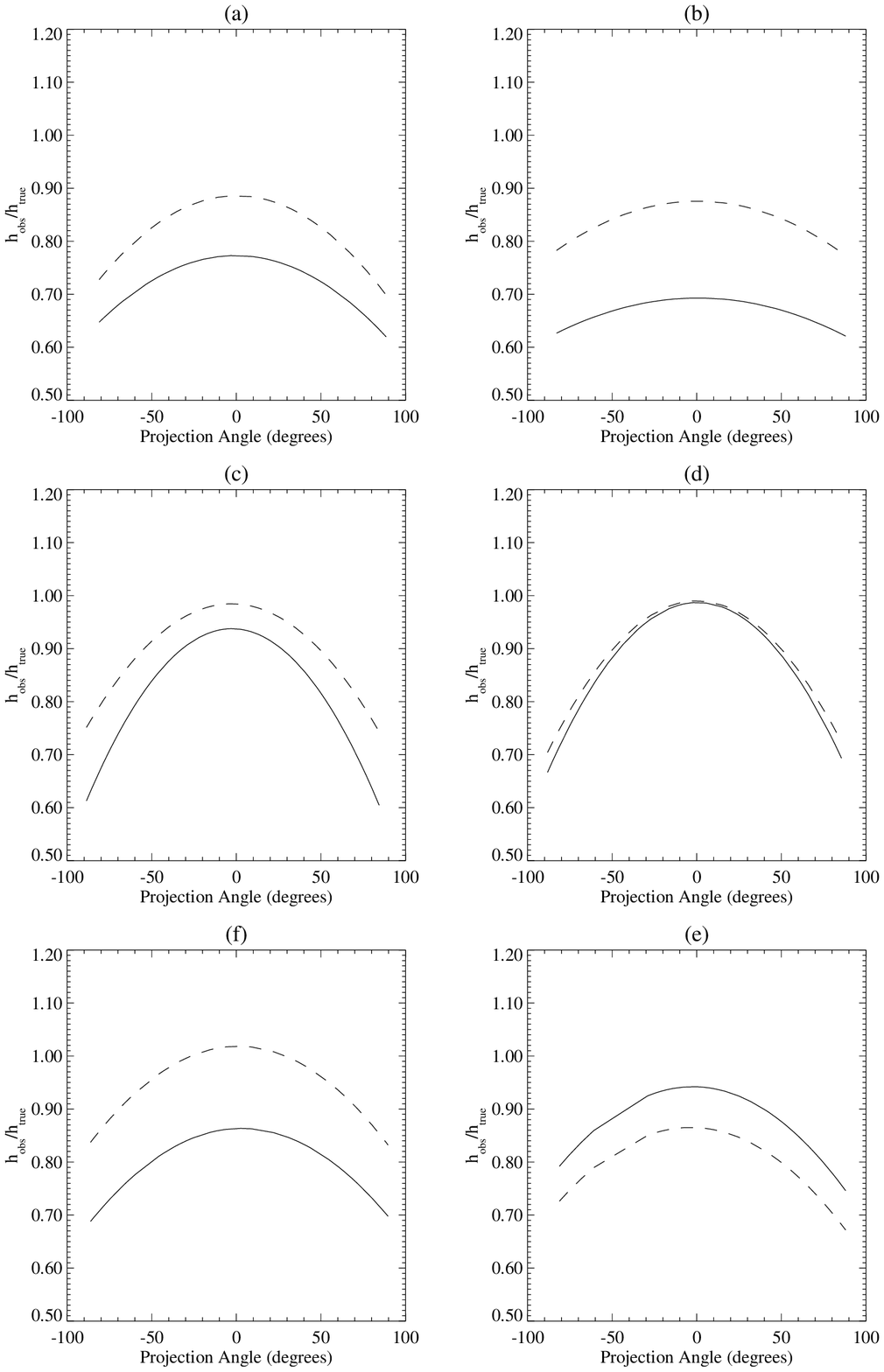}
\caption[]
{ }
\label{stat4a}
\end{figure}

\bigskip
{\bf Fig. \ref{stat4a}:} The ratio of the ``observed" to ``true" value of \Ho
defined by Eq. \ref{ratio} using observationally derived temperature {\it and} density 
information as a function of
projection angle along the merger axis (also cluster's major axis). The solid line represents
the use of an emission-weighted isothermal temperature. The dashed line represents the use
of a projected, emission-weighted temperature profile. As in Fig. \ref{stat4b}, the lines 
represent fits
to the distribution of randomly drawn LOS through the cluster core. Maximum residuals
to the fit are $\sim$0.1.  The projection angle is defined
as in Fig. \ref{stat4b}

\begin{figure}[htbp]
\centering \leavevmode
\epsfxsize=0.8\textwidth \epsfbox{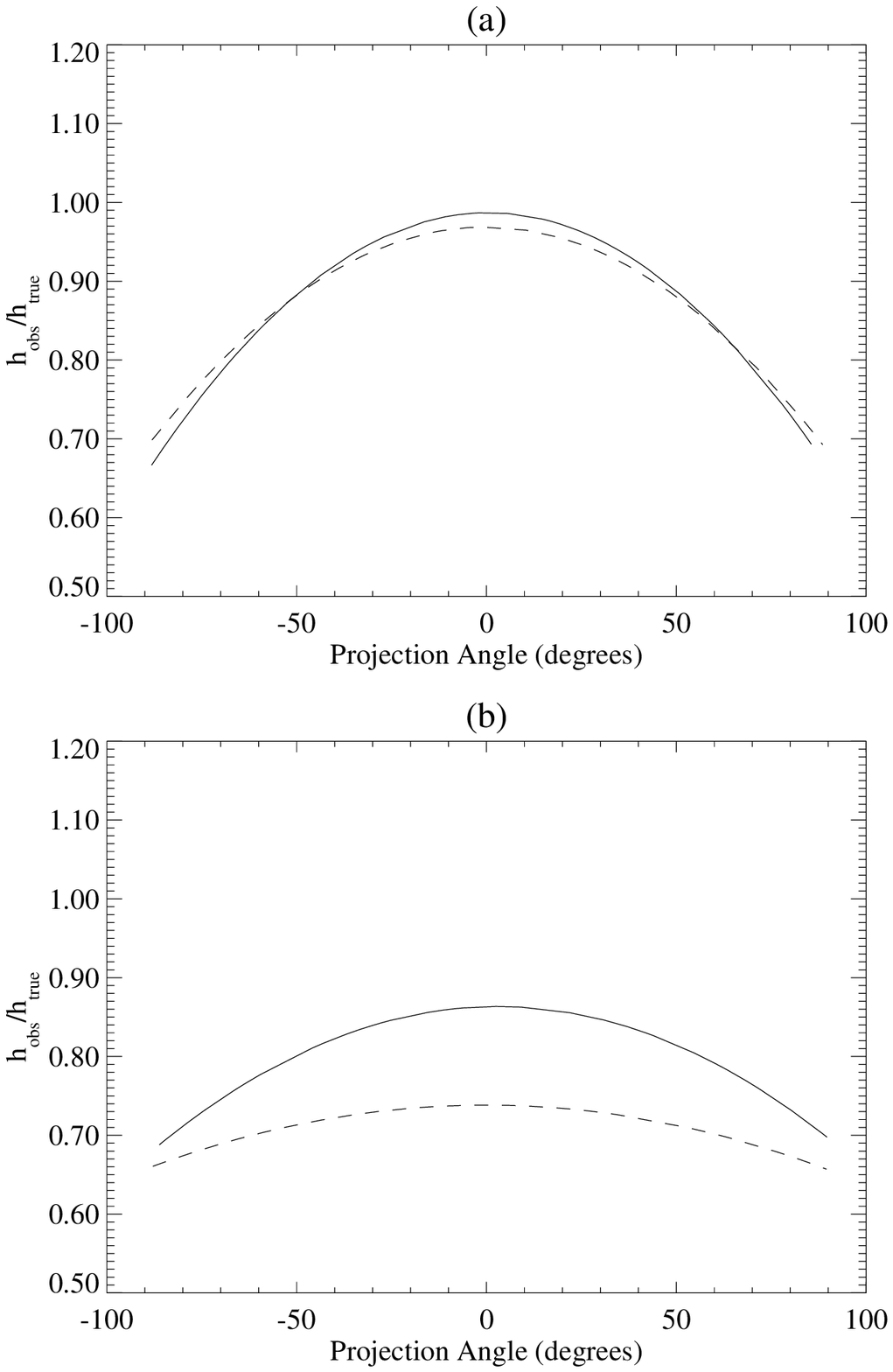}
\caption[]
{ }
\label{hevol}
\end{figure}

\bigskip
{\bf Fig. \ref{hevol}:}  Evolution of the ``observed" discrepancy in \Ho.
a) Evolution in Merger 4. Virtually no evolution is noted between 2.5 (solid)
and 5.0 (dashed) Gyrs. b) Evolution in merger 5. Here, we see the observed
discrepancy increase with time. As the merger remnant evolves, it becomes more
spherical and so the overestimate of \Ho at 0$^\circ$ is reduced, while the
underestimation of \Ho caused by the temperature estimate remains essentially the same.

\begin{figure}[htbp]
\centering \leavevmode
\epsfxsize=0.8\textwidth \epsfbox{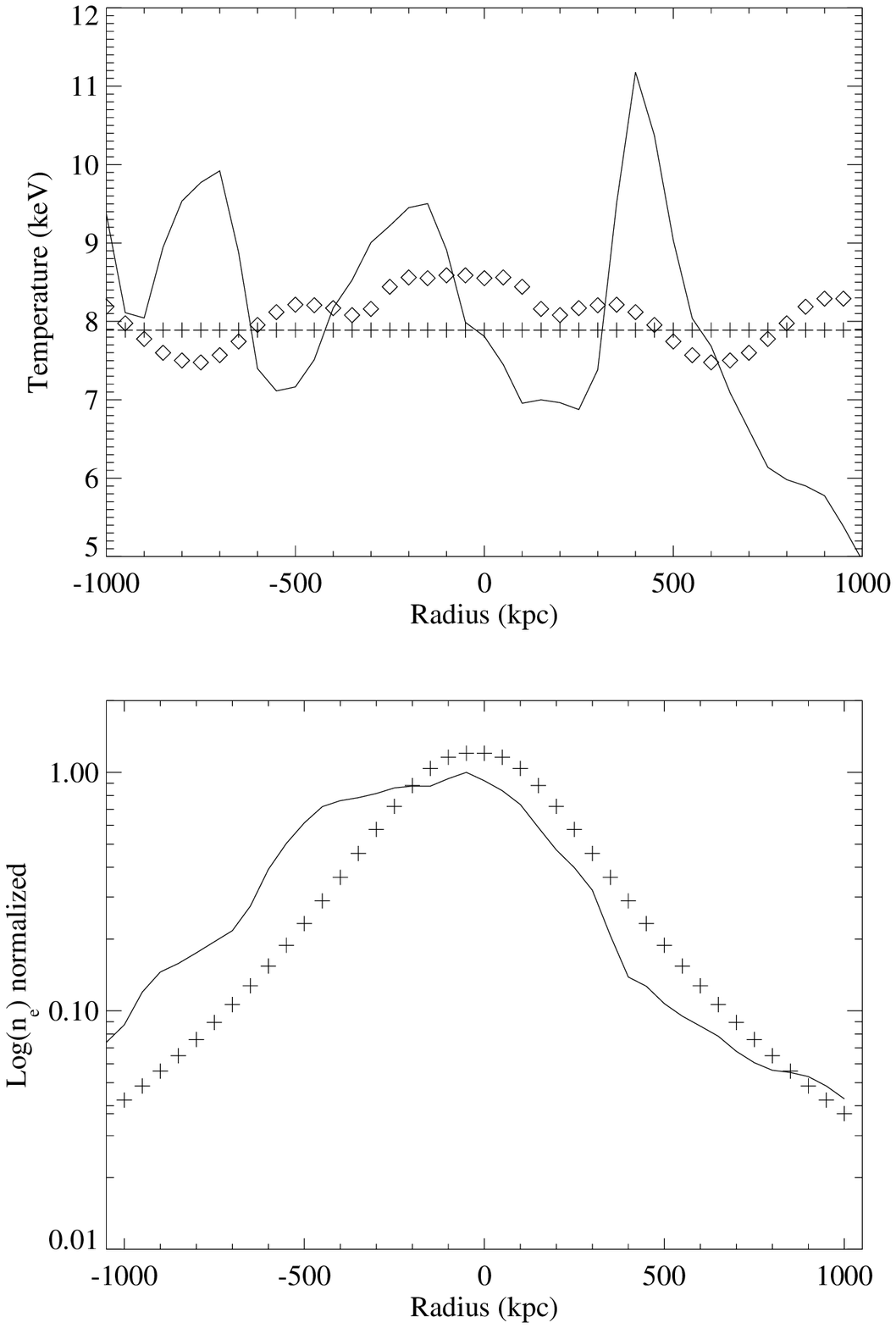}
\caption[]
{ }
\label{slice1}
\end{figure}

\bigskip
{\bf Fig. \ref{slice1}:} a) A comparison of the isothermal temperature (+)
(R$<$0.75 Mpc), the observed temperature profile ($\diamond$), and the
true temperature in a narrow beam along the merger axis (solid line) for
Merger 4 after 5 Gyrs. Note the observed temperatures are smoother and
tend to underestimate the true distribution.
b) A comparison of the observed density distribution (+) and the true
density along the merger axis (solid line). While the peak density is
slightly overestimated, the width of the 
observed distribution is severely underestimated.

\begin{figure}[htbp]
\centering \leavevmode
\epsfxsize=0.8\textwidth \epsfbox{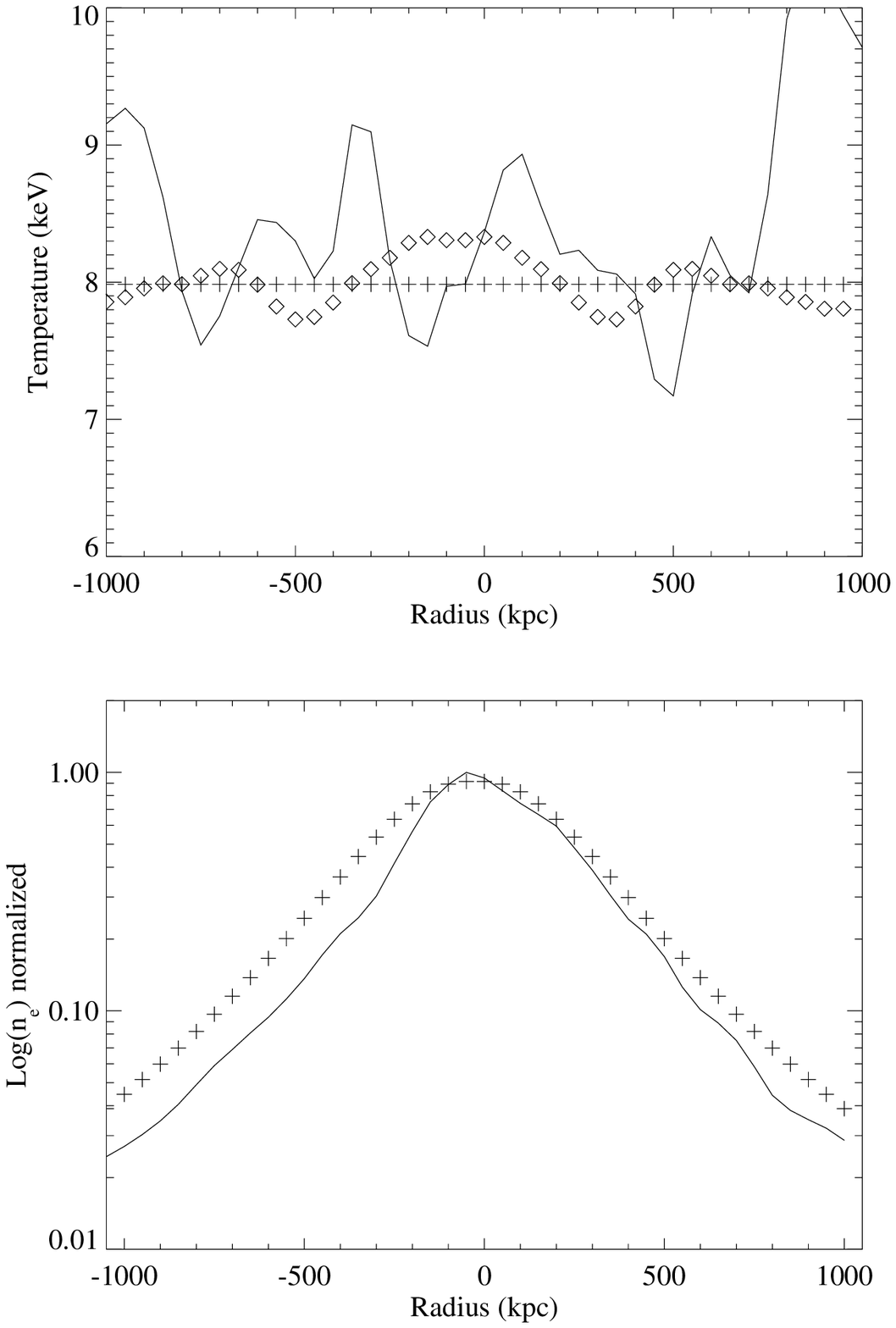}
\caption[]
{ }
\label{slice2}
\end{figure}

\bigskip
{\bf Fig. \ref{slice2}:} a) A comparison of the isothermal temperature (+)
(R$<$0.75 Mpc), the observed temperature profile ($\diamond$), and the
true temperature (solid line) in a narrow beam perpendicular to the merger axis (solid line) for
Merger 4 after 5 Gyrs. Similar to Figure \ref{slice1}, the observed temperatures are smoother and
tend to underestimate the true distribution.
b) A comparison of the observed density distribution (+) and the true
density (solid line) perpendicular to the merger axis (solid line). Contrary to Figure \ref{slice1},
the peak density is slightly underestimated, while the width of the distribution
is significantly overestimated.

\end{document}